\newif\ifDEBUG
\DEBUGtrue

\newif\ifARXIV
\ARXIVtrue

\ifARXIV
\documentclass[sigconf]{acmart}
\else
\documentclass[sigconf,authordraft]{acmart}
\fi
\AtBeginDocument{%
  }

\setcopyright{acmlicensed}
\copyrightyear{2018}
\acmYear{2018}
\acmDOI{XXXXXXX.XXXXXXX}
\ifARXIV
\acmConference[XXX '26]{ACM Conference on XXX}{2026}{United States}
\else
\acmConference[CCS '26]{ACM Conference on Computer and Communications Security}{Nov 15--19,
  2026}{The Hague, Netherlands}
\fi
\acmISBN{978-1-4503-XXXX-X/2018/06}

\usepackage{tikz}
\usepackage{amsmath}
\usepackage{comment}
\usepackage{listing}
\ifARXIV
\usepackage{minted}
\else
\usepackage{minted}
\fi
\usepackage[normalem]{ulem}

\usepackage[ruled,vlined,linesnumbered]{algorithm2e}
\DontPrintSemicolon
\SetKwProg{Fn}{Function}{}{end}
\SetKw{KwRet}{return}
\SetKw{KwAnd}{and}

\usepackage{xcolor}

\usepackage{xspace}
\usepackage{multirow} 
\usepackage{balance} 
\usepackage{fancyvrb} 
\usepackage{caption}
\usepackage{booktabs} 

\newcommand{\eg}{\textit{e.g.,}\xspace}
\newcommand{\etal}{\textit{et al.}\xspace}

\newif\ifSAVESPACE
\SAVESPACEfalse

\ifSAVESPACE

\else

\fi

\usepackage{todonotes}
\usepackage{soul}

\ifDEBUG
    \newcommand{\AH}[1]{\todo[color=cyan,inline]{AH:#1}}
    \newcommand{\AM}[1]{\todo[color=red,inline]{Machiry:#1}}
    \newcommand{\JD}[1]{\todo[color=yellow,inline]{JD:#1}}
    \newcommand{\TL}[1]{\todo[color=green,inline]{SA:#1}}
    \newcommand{\PA}[1]{\todo[color=orange,inline]{PA:#1}}
    \newcommand{\RC}[1]{\todo[color=cyan,inline]{RC:#1}}
    
    \newcommand{\KR}[1]{\todo[color=yellow,inline]{Kyle:#1}}
    \newcommand{\LS}[1]{\todo[color=green,inline]{LS:#1}}
    \newcommand{\HP}[1]{\todo[color=green,inline]{HP:#1}}
    \newcommand{\PP}[1]{\todo[color=lime,inline]{PP: #1}}

\else
    \newcommand{\AH}[1]{}
    \newcommand{\AM}[1]{}
    \newcommand{\JD}[1]{}
    \newcommand{\TL}[1]{}
    \newcommand{\PA}[1]{}
    \newcommand{\KR}[1]{}
    \newcommand{\LS}[1]{}
    \newcommand{\HP}[1]{}
    \newcommand{\PP}[1]{}
    \newcommand{\RC}[1]{}

\fi

\usepackage{xcolor}

\usepackage{hyperref}

\usepackage{cleveref}
\crefformat{section}{\S#2#1#3}
\crefname{figure}{Figure}{Figures}
\crefname{table}{Table}{Tables}
\crefname{theorem}{Theorem}{Theorems}
\crefname{thm}{Theorem}{Theorems}
\crefname{lemma}{Lemma}{Lemmata}
\crefname{equation}{Eqt.}{Eqts.}
\crefformat{Grammar}{Grammar #1}
\crefname{appendix}{Appendix}{Appendices}
\crefname{listing}{Listing}{Listings}
\crefname{algorithm}{Algorithm}{Algorithms}

\newcommand{\myparagraph}[1]{\paragraph{#1}}
\renewcommand{\myparagraph}[1]{\vspace{0.25em} \noindent \hspace{0.085cm}\underline{\textit{#1:}}\xspace}

\newcommand{\myinlineparagraph}[1]{\textbf{#1}\xspace}

\usepackage{pifont}

\usepackage{pifont}

\newcommand{\toolname}{AutoSOUP\xspace}
\newcommand{\msv}{memory-safety verification\xspace}
\newcommand{\Msv}{Memory-safety verification\xspace}
\newcommand{\MSV}{Memory-Safety Verification\xspace}

\begin{document}

\title{AutoSOUP: Safety-Oriented Unit Proof Generation for Component-level \MSV}


\author{Paschal Amusuo}
\affiliation{%
  \institution{Purdue University}
  \country{USA}}
\email{cpamusuo@purdue.edu}

\author{Ricardo Calvo}
\affiliation{%
  \institution{Purdue University}
  \country{USA}}
\email{rcalvome@purdue.edu}

\author{Dharun Anandayuvaraj}
\affiliation{%
  \institution{Purdue University}
  \country{USA}}
\email{dananday@purdue.edu}

\author{Taylor Le Lievre}
\affiliation{%
  \institution{Columbia University}
  \country{USA}}
\email{tjl2152@columbia.edu}

\author{Kevin Kolyakov}
\affiliation{%
  \institution{University of Waterloo}
  \country{Canada}}
\email{kkolyakov@uwaterloo.ca}

\author{Elijah Jorgensen}
\affiliation{%
  \institution{Purdue University}
  \country{USA}}
\email{jorgene1@purdue.edu}

\author{Aravind Machiry}
\affiliation{%
  \institution{Purdue University}
  \country{USA}}
\email{amachiry@purdue.edu}

\author{James C. Davis}
\affiliation{%
  \institution{Purdue University}
  \country{USA}}
\email{davisjam@purdue.edu}

\renewcommand{\shortauthors}{Amusuo et al.}

\begin{abstract}
Memory-safety errors remain a persistent source of zero-day vulnerabilities in low-level software. The problem is especially acute in embedded systems, where hardware protections are often limited and dynamic analysis is difficult to apply effectively. Memory-safety verification can provide stronger assurance by proving the absence of such errors or exposing violations when they exist. However, current verification workflows remain largely manual and require substantial specialized expertise, limiting their adoption in practice.

We present AutoSOUP, a system for automating component-level memory-safety verification through Safety-Oriented Unit Proofs. We formalize these unit proofs as artifacts that encode verification choices (scope, loop bounds, and environment models) for verifying safety properties, and introduce three techniques for deriving them automatically. To overcome the limitations of existing automation approaches, we further introduce LLM-As-Function-Call, a hybrid architecture that combines deterministic program synthesis with LLMs to automate these techniques and produce justifiable unit proofs. We evaluate AutoSOUP by assessing its ability to automate memory-safety verification and expose vulnerabilities in verified components, and we characterize the assumptions and guarantees of the resulting proofs.
\end{abstract}



\received{20 February 2007}
\received[revised]{12 March 2009}
\received[accepted]{5 June 2009}

\maketitle

\section{Introduction}

\begin{figure}
    \centering
    \includegraphics[width=\linewidth]{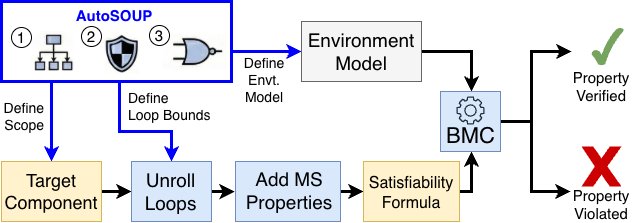}
    \caption{
    \toolname automatically identifies the functions to include in the verification scope, loop bounds, and environment model for which the resulting component-level \msv completes and provides useful guarantees of memory safety.
    }
    \label{fig:bmc-autosoup-overview}
\end{figure}

Memory-safety errors remain a persistent source of zero-day exploits in low-level software~\cite{alex_rebert_safer_2024, msrc_team_proactive_2019}.
These errors enable denial-of-service, data-theft, and remote-code-execution attacks~\cite{szekeres_sok_2013}.
Their risk is especially acute in embedded systems, which often lack standard hardware protections~\cite{abbasi_challenges_2019} and are difficult to analyze effectively with techniques such as fuzzing~\cite{muench_what_2018,ritvik2025lemix}.
After decades of recurring exploits and failures~\cite{miller_empirical_1990,szekeres_sok_2013,msrc_team_proactive_2019,crowdstrike_channel_file_291_2024}, government agencies and industry organizations increasingly advocate or mandate methods that provide stronger memory-safety guarantees~\cite{alex_rebert_securing_nodate, bob_lord_urgent_2023, noauthor_darpa_nodate,lw_eu_product_liability_2024}.

Formal \msv can provide such guarantees, but existing techniques remain difficult to apply at scale.
Deductive verification~\cite{filliatre_deductive_2011} can prove memory safety~\cite{klein_sel4_2009, li_memory_2021}, but it often requires specialized verification expertise, domain knowledge, and manual proof guidance~\cite{fonseca_empirical_2017, huang_lessons_2024}.
Bounded model checking (BMC)~\cite{iec61508-3,clarke_bounded_2001} offers a more automated alternative, but whole-program BMC does not scale to large software systems.
Recent industry case studies~\cite{chong_code-level_2020, wu_verifying_2024, wang_unsafecop_2025} favor a decomposition approach, wherein engineers verify components in isolation using \textit{``unit proofs''}~\cite{amusuo_unit_2025} that specify a component's verification scope, loop bounds, and environment models.
However, constructing these unit proofs remains challenging and error-prone:
  engineers must choose scopes, bounds, and models that are strong enough to expose genuine memory-safety errors, but constrained enough to keep verification tractable.
AWS's report~\cite{chong_code-level_2020} suggests a substantial engineering cost, with a verification rate of 1500 lines of code in one person-month.
No existing work automates these choices in a way that supports practical component-level memory-safety verification for real C software.

To address this gap, this paper introduces \toolname (\cref{fig:bmc-autosoup-overview}), a hybrid system for constructing unit proofs that enable useful \msv for embedded software.
The key idea is the notion of \emph{memory-safety-oriented unit proofs}: reusable verification artifacts that prioritize the scope, bounds, and environment models needed to preserve safety-relevant memory behavior, rather than faithfully model all execution behavior.
\toolname derives these choices using three techniques: resource-aware scope widening, property-guided loop-bound selection, and context-aware environment refinement.
To make this automation reliable and generalizable, \toolname uses a hybrid \emph{LLM-as-function-call} architecture~\cite{ji_artemis_2025, li_iris_2025}.
Rather than asking an LLM to synthesize an entire proof artifact, \toolname uses deterministic program-analysis workflows to drive the construction process and invokes LLMs for controlled tasks with explicit validation criteria.
As a result, each verification choice incorporated into the unit proof is tied to a specific algorithmic objective and validated before use.

We evaluate AutoSOUP on four embedded real-time operating systems.
AutoSOUP successfully produces valid unit proofs and verifies \(93\%\) of candidate targets, exposing \(66.7\%\) of evaluated vulnerabilities---\(38.7\%\) and \(28.5\%\) more than the next best-performing baselines.
Each of our techniques contributes to the final verification results by improving the derived bounds and environment models.
Moreover, AutoSOUP's unit proofs use simpler and more general environment assumptions than fidelity-oriented expert-written counterparts while achieving comparable verification outcomes.

In summary, this paper makes the following contribution:
\begin{itemize}
    \item We propose \emph{safety-oriented unit proofs}, a formulation of component-level \msv artifacts; and techniques for scope, bounds, and environment models.

    \item We operationalize the \emph{LLM-as-function-call} architecture for automating auditable and justifiable unit-proof construction.

    \item We design and implement \toolname to automatically construct safety-oriented unit proofs for C-language software.

    \item We evaluate \toolname on 177 components from four widely used embedded operating systems and report its utility, effectiveness, and cost for \msv.
\end{itemize}

\myparagraph{Significance}
\toolname is the first system to automate \msv through the construction of unit proofs that enable component-level bounded model checking.
Our results show that \toolname applies to substantial real-world codebases, provides useful memory-safety assurance, and exposes real security vulnerabilities.
Automated \msv can be practical enough to integrate into software-development workflows and help prevent memory-safety vulnerabilities before deployment.

\section{Background}
\label{sec:background}

\subsection{Memory-Safety and Verification}

\textit{Memory safety} means that all memory accesses performed by a program are valid under the semantics of the language~\cite{szekeres_sok_2013}.
Accesses must refer to memory that has been properly allocated, remain within the bounds of the target object, and occur during that object's lifetime~\cite{nagarakatte_full_2024, azevedodeamorimMeaningMemorySafety2018, hicksWhatMemorySafety2014}.
A memory safety error violates these conditions~\cite{nagarakatte_full_2024, vanoorschotMemoryErrorsMemory2023}, either spatially or temporally~\cite{szekeres_sok_2013}.
Violations of these conditions cause memory corruption and remain a major source of software failures, security vulnerabilities, and exploits~\cite{noauthor_project_nodate, msrc_team_proactive_2019, noauthor_channel_nodate, wee_here_nodate}.

\textit{\Msv} uses formal methods to establish that a program cannot perform invalid memory accesses, subject to explicit modeling assumptions.
Compared with testing and bug-finding techniques, its goal is not only to expose errors, but also to provide assurance that specified classes of memory violations are absent within the verified program.


Two approaches have been especially important for \msv.
\textit{Deductive verification} proves memory safety by reasoning over program behavior using specifications of program state, memory, and invariants~\cite{klein_sel4_2009, calcagno_compositional_2009}.
It can provide strong guarantees over all executions, but it requires substantial expertise and manual proof effort.
In contrast, \textit{Bounded Model Checking (BMC)}~\cite{ivancic_efficient_2008} checks memory-safety properties automatically by exploring executions within explicit bounds.

\begin{listing}
  \centering
  \caption{
  \textit{Left}: Program instrumented with memory-safety properties and its calling context.
  \textit{Right:} Unit proof with verification choices for the \texttt{process\_record} component.}
  \label{listing:sample-program-and-proof}

  \begin{minipage}[t]{0.48\columnwidth}%
\begin{minted}[
  fontsize=\scriptsize,
  linenos,
  frame=single,
  breaklines,
  xleftmargin=0.8em,
  numbersep=2pt,
  escapeinside=||,
  style=colorful
]{c}
size_t get_record_count() {
  // Complex logic returning [0,10]
  return count;
}

int handle_record(size_t i) {
  ...
}

void process_record(uint8_t *dst) {
  size_t n = get_record_count();
  // Bug!! should be i < n
  for (size_t i = 0; i <= n; ++i) {
    dst[i] = handle_record(i);
    |\textcolor{red}{\texttt{assert(isValidObject(dst));}}|
    |\textcolor{red}{\texttt{assert(ObjectSize(dst) > i);}}|
  }
}

void caller() {
  int dst[10];
  process_record(dst);
}
\end{minted}
  \end{minipage}%
  \hspace{0.006\columnwidth}%
  \begin{minipage}[t]{0.494\columnwidth}
\begin{minted}[
  fontsize=\scriptsize,
  linenos,
  frame=single,
  breaklines,
  xleftmargin=0.8em,
  numbersep=2pt,
  escapeinside=||,
  style=colorful
]{c}
Scope = {process_record, handle_record}.
Loop bound: {process_record.0: 11}

size_t get_record_count_m1() {
  uint8_t ret = nondet_int();
  |\textcolor{blue}{\texttt{assume(ret < 10);}}|
  return ret;
}

size_t get_record_count_m2() {
  uint8_t ret = nondet_int();
  |\textcolor{blue}{\texttt{assume(ret <= 10);}}|
  return ret;
}

void harness(void) {
  uint8_t dst_size = nondet_int();
  uint8_t *dst = malloc(dst_size);
  |\textcolor{blue}{\texttt{assume(dst != NULL);}}|
  process_record(dst);
}
\end{minted}
  \end{minipage}
\end{listing}

\subsection{Bounded Model Checking and Unit Proofing}
\label{sec:bg-bmc-up}

Tools for Bounded Model Checking (BMC)~\cite{kroening_cbmc_2014, kani_getting_nodate} specify program properties as Boolean assertions over variable states, automatically check these assertions using constraint-solving tools, and report counterexample traces~\cite{martin_brain_cbmc_nodate}.
For memory-safety verification, BMC instruments the program with memory-safety assertions, unrolls loops and recursion, translates the resulting bounded program into a satisfiability formula, and applies constraint solvers.
If no violating trace exists, BMC proves that the instrumented properties hold.
If a violation exists, BMC returns a counterexample. 

\Cref{listing:sample-program-and-proof} illustrates how BMC can be used to verify a component's memory safety.
An engineer verifies a connected subset of a program to keep verification tractable (\textit{verification scope}); for example, they may verify \texttt{process\_record} and \texttt{handle\_record} while excluding the \texttt{get\_record\_count} callee.
The engineer must also choose the maximum number of times to unroll loops (\textit{loop bounds}), because BMC operates only on loop-free programs, and provide assumptions about the behavior of callers and callees outside the chosen scope (\textit{environment model}).

These choices---verification scope, loop bounds, and environment model---are encoded in a \textit{unit proof}, shown on the right of \cref{listing:sample-program-and-proof}.
For example, Program Lines~15--16 assert that \texttt{dst} points to a valid allocated object and that the object is large enough for the indexed write; the unit proof then determines the scope, bounds, and assumptions under which BMC checks those assertions.

\subsection{Verification Choice Fidelity}
\label{subsec:bg-verif-fidelity}

The guarantees afforded by BMC depend on the choices encoded in a unit proof.
For example, on Line 13 of~\cref{listing:sample-program-and-proof}, setting the bound for the program loop to 1 may suffice to check that \texttt{dst} is valid, but not that \textit{all} indexed accesses are valid.
Similarly, the overly-constrained model \texttt{get\_record\_count\_m1} (Unit Proof Lines~4--8) masks the violation, 
whereas \texttt{get\_record\_count\_m2} exposes it.

We define \textit{verification choice fidelity (VCF)} as the degree to which these choices reflect the behavior of the real system.
Formally, we characterize fidelity as \(VCF = \langle S_{cf}, B_f, E_f \rangle\), where \(S_{cf}\) denotes verification-scope fidelity, \(B_f\) denotes loop-bound fidelity, and \(E_f\) denotes environment-model fidelity.
This notion is related to model fidelity in systems modeling~\cite{mashkoor_validation_2021} and execution fidelity in firmware rehosting~\cite{ritvik2025lemix,feng_p2im_2020} but focuses on choices that affect bounded model checking.
Higher-fidelity choices preserve implementation behavior and can support stronger guarantees.
Lower-fidelity choices simplify semantics, reducing development cost.

Many existing approaches to component-level BMC are \textit{fidelity-oriented}. 
They rely on experts to encode detailed knowledge of the system~\cite{chong_code-level_2020}, infer artifacts from specifications~\cite{wu_verifying_2024} or unit tests~\cite{wang_unsafecop_2025}, or refine artifacts manually using verification results~\cite{amusuo_unit_2025-1}.
Such proofs can provide strong guarantees, but require deep knowledge of the component and its role in the surrounding program.
It is therefore hard to create them automatically. 

Other approaches rely on automatable choices, such as fixed function-level scopes~\cite{cho_blitz_2013, ivancic_dc2_2011}, uniform loop bounds~\cite{takhar_memory-safety_2025}, or environment models inferred by program analysis~\cite{ivancic_dc2_2011, calcagno_compositional_2009}.
However, these methods focus on restricted property sets, limiting their use for \msv; or require expert guidance to apply them in real software.
This gap motivates methods that derive verification choices automatically while preserving the safety-relevant behavior needed for useful \msv guarantees.

\section{Problem Statement}
\label{subsec:ps-problem}



Our goal is to automate component-level memory-safety verification through the creation of unit proofs.
This requires identifying verification choices that provide meaningful memory-safety guarantees while keeping verification tractable.

We formalize the problem as follows.
Given a software system~\(S\), a component entry point~\(C_e\), a set of memory-safety properties~\(Q\), and a resource budget~\(R\), automatically construct a unit proof~\(U(V)\) that encodes the verification choices
\[
    V = (S_c, B, E),
\]
where \(S_c \subseteq S\) is a connected set of functions rooted in \(C_e\), \(B\) maps each loop reachable in \(S_c\) to a maximum unrolling bound, and \(E\) is an environment model that defines assumptions for functions outside \(S_c\).
The resulting proof should allow BMC to verify \(Q\) conclusively within \(R\),
while ensuring the result reflects the memory safety of the verified functions \(S_c\) in the original system \(S\).
In \cref{listing:sample-program-and-proof}, this means constructing the unit proof on the right so that BMC exposes the memory-safety error on the left.

No existing work solves this problem automatically.
Prior work on \msv either requires experts to define these choices~\cite{chong_code-level_2020, wu_verifying_2024, wang_unsafecop_2025}, or uses fixed scope, loop bounds and program-analysis-inferred environment models~\cite{cho_blitz_2013, ivancic_dc2_2011, calcagno_compositional_2009} that require expert adaptation to real software.
In this work, we automatically derive verification choices and construct executable unit proofs for real C software.

\subsection{Success Criteria}
\label{subsec:ps-success-criteria}

A unit proof should support useful memory-safety guarantees.
Building on~\cite{amusuo_unit_2025-1},
  we distinguish five requirements:

\myparagraph{(1) Structural validity}
It should compile and verify the intended component identified by the entry point \(C_e\).
For example, the unit proof in \cref{listing:sample-program-and-proof} must verify \texttt{process\_record}.

\myparagraph{(2) Conclusiveness~\cite{amusuo_unit_2025-1}}
It should produce a verification result within the resource budget \(R\).
The result may be either a proof that the target memory-safety properties \(Q\) hold, or a counterexample showing the states and execution trace that violate them.

\myparagraph{(3) Verification coverage~\cite{amusuo_unit_2025-1}}
It should verify as much reachable code in its verification scope as possible.
We define \textit{verification coverage} as the proportion of included lines that are exercised and checked by the unit proof.
This metric is analogous to line coverage in unit testing and fuzzing. 

\myparagraph{(4) Result validity} It should make verification choices that neither encode spurious violations nor mask feasible memory-safety violations within the checked scope, bounds, and assumptions.
Reported violations should correspond to real memory-safety errors.

\myparagraph{(5) Maintainability}
Unit proofs are reusable and auditable artifacts.
As a result, their format should be familiar to software engineers so that the verification choices can be inspected and refined and the unit proofs maintained as the code evolves.

\vspace{0.1cm}
\noindent
Automatically generating unit proofs that satisfy these requirements is non-trivial because the requirements can conflict.
For example, a system may improve conclusiveness with shallow bounds or simple models, but reduce verification coverage or result validity.
Conversely, techniques optimized for result validity may compromise coverage and tractability, or yield unit proofs too complex for engineering teams.
Prior work~\cite{amusuo_unit_2025-1} introduced conclusiveness and verification coverage to assess unit proof completeness.
We add structural validity, result validity, and maintainability to capture risks introduced by automation: a generated unit proof may verify the wrong entry point, encode choices that produce invalid results, or become too complex for maintainers to inspect and refine.

\subsection{System and Threat Model}
\label{subsec:system-threat-model}

\myparagraph{System model}
We consider software systems \(S\) written in the C language, which is often used for embedded software development.
We focus on components that process untrusted inputs from external interfaces, \eg network channels, since these interfaces form primary attack surfaces.
We exclude programs in which memory references can be shared across processes (threads or tasks) and that are prone to race conditions, because they violate the linear execution assumptions of standard bounded model checking tools. 

\myparagraph{Threat model}
We consider attackers who can supply inputs through external interfaces and thereby trigger violations of the target memory-safety properties \(Q\).
Such inputs may flow directly to memory-safety sinks or indirectly influence control/data flow so that the component reaches an unsafe memory state.

\section{AutoSOUP Design and Implementation}
\label{sec:autosoup-design}
We designed \toolname to automate component-level \msv through the creation of unit proofs (\cref{fig:autoup-design}).

\ifARXIV

\begin{figure*}
    \centering
    \includegraphics[width=0.9\linewidth, trim=0 0 0.3cm 0, clip]{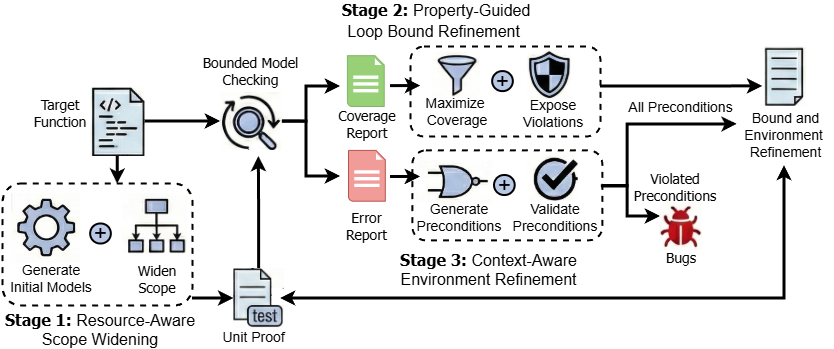}
    \caption{
    \toolname system diagram.
    \toolname connects safety-oriented unit-proof construction with LLM-as-function-call automation.
    Across all stages, deterministic workflows derive the verification choices \(V=(S_c,B,E)\), delegate bounded semantic tasks to LLM agents, and validate returned artifacts before incorporating them into the unit proof.
    }
    \label{fig:autoup-design}
\end{figure*}

\else

\begin{figure*}
    \centering
    \includegraphics[width=0.62\linewidth, trim=0 0 0.3cm 0, clip]{figures/autosoup-design.drawio.png}
    \caption{
    \toolname system diagram.
    \toolname connects safety-oriented unit-proof construction with LLM-as-function-call automation.
    Across all stages, deterministic workflows derive the verification choices \(V=(S_c,B,E)\), delegate bounded semantic tasks to LLM agents, and validate returned artifacts before incorporating them into the unit proof.
    }
    \label{fig:autoup-design}
\end{figure*}

\fi

\subsection{Key Ideas}
\toolname explores two ideas: safety-oriented derivation of verification choices (\cref{subsubsec:design-soup-idea}) and LLM-as-function-call automation (\cref{subsubsec:design-automation}).

\subsubsection{Memory-Safety-Oriented Unit Proofs}
\label{subsubsec:design-soup-idea}

We introduce \textit{memory-safety-oriented unit proofs}: unit proofs whose verification choices \(V=(S_c,B,E)\) are tailored to the instrumented memory-safety properties \(Q\).
Unlike fidelity-oriented unit proofs, which choose \(V\) to approximate the program's behavior, memory-safety-oriented unit proofs choose \(V\) to preserve the behavior needed to prove or refute \(Q\) within the resource budget \(R\).
This gives a safety-oriented interpretation of each choice:
\(S_c\) should include code that contributes checkable memory-safety behavior;
\(B\) should be large enough to cover or expose property-relevant loop behavior;
and \(E\) should expose possible violations while excluding infeasible ones.

\toolname realizes this interpretation through three techniques.

\myparagraph{1. Resource-aware scope widening}
Prior fidelity-oriented approaches rely on experts to select verification scopes~\cite{chong_code-level_2020, takhar_memory-safety_2025}.
Automating this selection is difficult because it requires reasoning about both inter-function semantics and verification cost.
However, prior work~\cite{amusuo_unit_2025-1} suggests that memory-safety verification often does not require preserving all inter-function behavior.

We therefore treat scope selection as a resource-aware coverage problem rather than a full semantic-modeling problem.
\toolname widens \(S_c\) incrementally, using simple models for functions that remain outside the current scope.
This lets the verifier check memory-safety properties in the included code without requiring precise models for every external callee.
A widened scope is useful only if it increases the code and properties checked while remaining within the resource budget \(R\).
This reframing avoids the need to predict the globally best scope in advance: \toolname starts with a small, tractable scope and expands only as needed.

\myparagraph{2. Property-guided loop-bound refinement}
Prior fidelity-oriented approaches rely on experts to define loop bounds~\cite{chong_code-level_2020} or loop invariants~\cite{gadelha_handling_2017}.
Other approaches use uniform bounds for all loops~\cite{takhar_memory-safety_2025}, which can be expensive and still miss memory-safety-relevant behavior.
We instead treat loop-bound selection as a property-guided refinement problem.
The intuition is that many memory-safety properties depend on local program states and are affected most directly by nearby loops.
Thus, \toolname refines \(B\) only for loops whose iterations affect the reachability or violation of instrumented memory-safety properties (\eg the loop on Line 13 of~\cref{listing:sample-program-and-proof}).

\myparagraph{3. Context-aware environment refinement}
Environment models create a different tension: permissive models expose possible memory-safety violations, but can also produce infeasible counterexamples; overly constrained models avoid spurious reports, but can mask real violations.
We therefore extend counterexample-guided environment refinement~\cite{ivancic_dc2_2011} with context validation.
\toolname starts with permissive models that expose possible violations, infers preconditions that suppress infeasible violations, and then validates those preconditions against calling contexts in the original program.
Preconditions satisfied by the surrounding program become explicit assumptions in \(E\); violated preconditions identify feasible memory-safety errors.
In \cref{listing:sample-program-and-proof}, this process distinguishes assumptions needed for \texttt{process\_record} to be memory safe from assumptions violated by the actual implementation of \texttt{get\_record\_count()}.

\vspace{0.1cm}
\textbf{Together}, these techniques derive \(S_c\), \(B\), and \(E\) in a property-directed way.
The resulting unit proofs \(U(V)\) are safety-oriented because they preserve behavior relevant to the target safety properties rather than all program behavior.
Although we focus on memory safety, the same formulation applies to safety properties expressible as Boolean predicates over program states.

\subsubsection{Automating Safety-Oriented Unit Proofs}
\label{subsubsec:design-automation}

Automating safety-oriented unit proofs requires both project-specific reasoning and auditable verification choices.
Program analysis can enforce syntactic and semantic constraints, but struggles with build configuration, local code idioms, and scalable semantic inference.
LLMs can handle such project-specific reasoning, but unconstrained generation is unsuitable because invalid scopes, bounds, or environment assumptions can invalidate the resulting verification guarantees.

\toolname balances reliability and generalizability with an \textit{LLM-as-function-call} architecture.
Deterministic algorithms control the proof-construction workflow and define the objective of each step.
When a step requires semantic code understanding or project-specific adaptation, \toolname delegates a bounded task to a tool-equipped LLM agent.
Program-analysis modules then validate the returned artifact before it is incorporated into the unit proof.

\subsection{Resource-Aware Scope Widening}
\label{subsec:resource-aware-scope-widening}

Resource-aware scope widening derives \(S_c\) by incrementally adding semantically related code that may contain checkable memory-safety properties, retaining expansions only while verification remains within the resource budget \(R\).
We use source files as the unit of expansion because related functions are often colocated, making file-level widening a coarse but useful approximation of semantic locality.
The algorithm proceeds in three steps (cf.~\cref{app:subsec:resource-aware-scope-widening}).

\myparagraph{Step 1: Initialize the scope, bounds, and input model}
AutoSOUP initializes \(S_c\) with the file containing \(C_e\), sets all loop bounds to 1, and constructs a type-directed input model for \(C_e\): primitive arguments receive unconstrained symbolic values, while pointer arguments receive valid allocated objects containing unconstrained values.
An LLM agent synthesizes this input model and recovers the build configuration needed to compile the entry point's parent file, including headers, include paths, macros, and mandatory project configurations.
AutoSOUP accepts the result only if deterministic checks confirm that the proof compiles, calls \(C_e\), and introduces no preconditions beyond the intended type-directed initialization.

\myparagraph{Step 2: Model external calls}
AutoSOUP identifies call edges that cross the current scope boundary and replaces their targets with simple type-based models.
Primitive returns are modeled as unconstrained symbolic values, while pointer returns are modeled as valid allocated objects to avoid irrelevant invalid-pointer states that inflate the solver state space.
An LLM agent synthesizes and integrates these models from the recovered call graph and return types; deterministic checks confirm that the resulting unit proof remains structurally valid.

\myparagraph{Step 3: Widen the scope}
AutoSOUP checks the provisional unit proof against the configured time, memory, and file-depth budgets.
If verification remains within \(R\), it widens \(S_c\) by adding files that define previously modeled callees, selecting among ambiguous definitions by longest common path prefix to the in-scope caller.
An LLM agent recovers build configuration for newly added files, and AutoSOUP repeats external-call modeling and scope widening until verification exceeds \(R\) or no additional files can be added.

\setcounter{AlgoLine}{0}
\begin{algorithm}
\caption{Property-guided loop-bound and model refinement.
Given a verification instance with initial loop bounds, refine the bounds, build configuration, and environment models to cover property-relevant code and expose loop-dependent memory-safety violations.}
\label{alg:property-guided-loop-bound-refinement}
\small

\KwIn{Verification scope $S_c$, properties $Q$, loop bounds $B$, environment model $E$, resource budget $R$}
\KwOut{Refined loop-bound map $B$, refined environment model $E$}

\Fn{\texttt{PropertyGuidedRefinement}$(S_c, Q, B, E, R)$}{
    \tcp{Step 1: Cover property-relevant code}
    $G \leftarrow \texttt{UncoveredPropertyBlocks}(S_c, Q, B, E)$\;
    \ForEach{$g \in G$}{
        $\rho \leftarrow \texttt{ClassifyCoverageGap}(g, S_c, B, E)$\;
        $(B', E') \leftarrow \texttt{RepairCoverageGap}(g, \rho, B, E)$\;

        \If{\texttt{ValidCoverageRefinement}$(S_c, Q, B', E', R)$}{
            $(B, E) \leftarrow (B', E')$\;
        }
    }

    \tcp{Step 2: Expose loop-dependent property violations}
    $L \leftarrow \texttt{LoopsWithIncompleteUnwinding}(S_c, B, E)$\;
    \ForEach{$\ell \in L$}{
        $P_\ell \leftarrow \texttt{LoopDependentProperties}(\ell, Q)$\;
        \If{$P_\ell \neq \emptyset$}{
            $n \leftarrow \texttt{MinBoundToViolate}(\ell, P_\ell)$\;
            $B' \leftarrow B$; $B'[\ell] \leftarrow \max(B[\ell], n)$\;

            \If{\texttt{ValidBoundRefinement}$(S_c, Q, B', E, R)$}{
                $B \leftarrow B'$\;
            }
        }
    }

    \KwRet $(B, E)$\;
}

\end{algorithm}

\subsection{Property-Guided Loop \& Model Refinement}
\label{subsec:property-guided-bound-refinement}

Property-guided refinement derives the bounds~\(B\) and model refinements~\(E\) to exercise memory-safety-relevant behavior in~\(S_c\).
It uses the instrumented properties~\(Q\) as the oracle for deciding which refinements matter.
It honors the resource budget $R$ by increasing bounds or refining models only to cover property-relevant code or expose violations.
Our technique has two steps (\cref{alg:property-guided-loop-bound-refinement}).

\myparagraph{Step 1: Cover property-relevant code}
A violation of \(q \in Q\) can be exposed only if the line containing \(q\) is reached and checked.
Coverage can be blocked by three factors: insufficient loop bounds, missing compile-time configuration, or external-call side effects not captured by return-value-only models.

\toolname runs BMC in coverage mode, identifies uncovered property blocks, and uses an LLM agent to classify the blocking factor.
It then applies the corresponding local repair: increase the relevant loop bound, adjust the configuration, or refine the external model to write unconstrained symbolic values through affected pointer arguments.
The taxonomy and repair rules constrain the agent so that refinements remain local and aligned with the technique.
A refinement is accepted only if the unit proof remains semantically valid, the target block becomes covered, and overall verification coverage does not decrease.

\myparagraph{Step 2: Expose loop-dependent property violations}
Covering a property does not imply that the current loop bounds can expose its violation.
\toolname inspects loops from the coverage report whose current bounds were insufficient for complete unwinding.
For each such loop, it uses an LLM agent to determine whether the loop can affect violation of a nearby memory-safety property and, if so, to estimate the minimum bound likely to expose a violation.
The prompt constrains this estimation using local program semantics, such as memory-region size, access stride, allocation constraints, and loop guards.
A proposed bound is accepted only if it modifies the intended loop bound and preserves structural validity, conclusiveness, and verification coverage.
If the bound causes verification to exceed \(R\), \toolname reports the bound but does not apply it.

\subsection{Context-Aware Env. Model Refinement}
\label{subsec:context-aware-env-refinement}

Property-guided refinement in~\cref{subsec:property-guided-bound-refinement} maximizes exposure of violations of \(Q\) using permissive models \(E\).
However, violations found under permissive models may be infeasible in the broader system~\(S\).
For example, the assertion on Program Line~15 in \cref{listing:sample-program-and-proof} is violated if the input model provides a null pointer, even if the actual caller provides a statically allocated array.
Our context-aware environment refinement separates infeasible violations (caused by overly permissive environment assumptions) from genuine memory-safety errors.
This technique operates in two steps (\cref{alg:context-aware-env-refinement}).

\setcounter{AlgoLine}{0}
\begin{algorithm}
\caption{Context-aware environment model refinement.
The algorithm infers approximate preconditions for violated memory-safety properties, validates them against calling contexts, and reports caller-feasible violations as memory-safety errors.}
\label{alg:context-aware-env-refinement}
\small

\KwIn{Software system $S$, component entry point $C_e$, Environment model $E$, Property violations $Q_v$}
\KwOut{Refined environment model $E$, memory-safety error set $\mathcal{M}$}

\Fn{\texttt{ContextAwareEnvRefinement}$(S, C_e, E, Q_v)$}{
    $\mathcal{M} \leftarrow \emptyset$\;
    $\mathcal{W} \leftarrow \texttt{ParseViolationReport}(Q_v)$\;
    \tcp{$\mathcal{W}$ contains tuples $(q,w(q))$}

    \ForEach{$(q,w(q)) \in \mathcal{W}$}{
        $\phi \leftarrow \texttt{InferApproxPrecondition}(E,q,w(q))$\;
        $(\phi', \mathcal{B}) \leftarrow \texttt{ValidatePrecondition}(S, C_e, q, \phi)$\;

        $E \leftarrow E \cup \{\phi'\}$\;
        $\mathcal{M} \leftarrow \mathcal{M} \cup \mathcal{B}$\;
    }

    \KwRet $(E,\mathcal{M})$\;
}

\BlankLine

\Fn{\texttt{ValidatePrecondition}$(S, C_e, q, \phi)$}{
    $\mathcal{B} \leftarrow \emptyset$;
    $\phi' \leftarrow \phi$\;
    $\mathcal{C} \leftarrow \texttt{CallsitesOf}(C_e, S)$\;

    \ForEach{$c \in \mathcal{C}$}{
        $\psi_c \leftarrow \texttt{PathConstraints}(c, S)$\;

        \If{$\psi_c \not\models \phi'$}{
            $\hat{\phi} \leftarrow \texttt{WeakenPrecondition}(\phi',\psi_c)$\;

            \uIf{$\texttt{SatisfiesProperty}(\hat{\phi},q)$}{
                $\phi' \leftarrow \hat{\phi}$\;
            }
            \Else{
                $\mathcal{B} \leftarrow \mathcal{B} \cup \{(q,\psi_c,\phi')\}$\;
            }
        }
    }

    \KwRet $(\phi',\mathcal{B})$\;
}
\end{algorithm}

\myparagraph{Step 1: Infer approximate weakest preconditions}
Following counter-example-guided environment refinement, \toolname infers preconditions that suppress reported violations of memory-safety properties.
These preconditions need not be logically weakest, since weakest-precondition inference is often computationally expensive~\cite{cho_blitz_2013}.
Instead, they should be weak enough to preserve safe states while strong enough to suppress the target violation.

\toolname parses the verification report to extract each violated property \(q \in Q\), its location \(\mathit{loc}(q)\), and its counterexample witness \(w(q)\).
For each tuple \((q,\mathit{loc}(q),w(q))\), an LLM agent infers a precondition that keeps \(\mathit{loc}(q)\) covered but suppresses the violation of \(q\).
The prompt guides the agent to identify the violated condition, propagate it backward through local dataflow and path constraints, and stop at the external model or input responsible for the value.
The agent can inspect witnesses, navigate code, and test candidate preconditions.
A candidate is accepted only if it suppresses the target violation without reducing structural validity, conclusiveness, coverage, or the number of checked properties.

\myparagraph{Step 2: Validate and refine against calling contexts}
An inferred precondition may exclude feasible caller states that would not violate \(q\), or it may reveal that the surrounding program can trigger the violation.
\toolname therefore validates each accepted precondition against calling contexts in \(S\).
Using a pre-indexed call graph, it identifies callsites of \(C_e\) and implementations of modeled functions.
For each context, an LLM agent identifies path constraints reaching the callsite or implementation and checks whether those constraints imply the inferred precondition.

Validation has three outcomes.
If a calling path violates the precondition but still satisfies \(q\), the agent weakens and revalidates the precondition.
If the precondition holds across the checked contexts, it becomes an explicit assumption in \(E\).
And of course, if a calling path violates the precondition and triggers \(q\), \toolname reports the path as a feasible memory-safety error in \(S\).
These assumptions make the generated unit proof auditable: the component is verified only under the preconditions recorded in its environment model.

\subsection{Guarantees and Limitations}

We discuss the guarantees and limitations of \toolname.

\subsubsection{Guarantees.}
Unit proofs generated by \toolname provide bounded formal guarantees of memory safety, backed by BMC.
They establish that \textit{no execution from the component entry point violates a verified memory-safety property, provided the execution stays within the scope, loop bounds, and assumptions encoded in the unit proof}.
This is the same form of guarantee provided by expert-written unit proofs~\cite{chong_code-level_2020}; its strength depends on the correctness and completeness of the encoded verification choices.
\toolname strengthens these choices by expanding safety-relevant scope, increasing bounds needed to expose property-relevant behavior, and making environment assumptions explicit.

\subsubsection{Limitations.}
Three limitations constrain these guarantees.
First, \toolname uses LLMs for project-specific reasoning.
To reduce unreliable outputs, deterministic workflows issue bounded tasks and validate outputs before incorporating them into the unit proof.
As program analyses improve, these modules can be replaced with deterministic techniques.
Second, \toolname relies on static analysis to recover call graphs for scope widening and precondition validation.
Imprecision, especially around indirect calls, may cause \toolname to miss relevant call edges; engineers can reduce this risk by providing accurate application call graphs.
Third, \toolname widens scope at file granularity.
Large files may exceed \(R\) and prevent wider scopes from being explored.

\subsection{Implementation}
\label{subsec:implementation}

We implement AutoSOUP in 15463 lines of Python and 1230 lines of prompts.
Resource-aware scope widening uses 1227 lines of Python and 292 prompt lines; property-guided loop-bound refinement uses 775 lines of Python and 316 prompt lines; and context-aware environment refinement uses 1091 lines of Python and 479 prompt lines.

\myparagraph{Bounded model checking}
AutoSOUP uses the ANSI-C Bounded Model Checker (CBMC)~\cite{kroening_cbmc_2014} as its verification backend.
AutoSOUP can be used with most bounded model checkers with memory-safety instrumentation, as elaborated in~\cref{app-sec:memory-safety-properties}. 

\myparagraph{Program-analysis modules}
AutoSOUP integrates components for deterministic program analysis. 
We use CBMC's \texttt{goto-instrument} to extract call-graph and symbol information from the compiled verification scope, which lets AutoSOUP check that the unit proof invokes the target entry point and identify undefined external callees during scope widening.
We use libclang to locate function-pointer calls in the verification scope, allowing AutoSOUP to constrain them to selected models or concrete targets.
We use cscope to index the software system \(S\), recover project-level call relations, locate candidate function definitions, and identify calling contexts for validating inferred preconditions.

\myparagraph{LLM-driven modules}
AutoSOUP implements its LLM-driven modules using the OpenAI Python SDK~\cite{noauthor_openaiopenai-python_2026} and LiteLLM~\cite{noauthor_litellm_nodate}.
The OpenAI SDK provides access to the GPT-family models used in our evaluation, while LiteLLM supports additional providers, including open-source and self-hosted models.

We expose three tools to the agents:
\textit{(i)} a \textit{containerized terminal tool} lets agents inspect source files, build logs, and verification reports;
\textit{(ii)} a \textit{cscope-based navigation tool} supports code search, definition lookup, and call-graph queries;
and
\textit{(iii)} a \textit{unit-proof validation tool} analyzes the compiled unit and verification report after each proposed change to confirm the proof compiles, calls the target entry point, satisfies the requested refinement task, and does not reduce line coverage or the number of covered or verified properties.

\section{Evaluation}
\label{sec:evaluation}

We structure our evaluation with four research questions:

\begin{itemize}
    \item \textbf{RQ1:} Can \toolname generate useful unit proofs for \msv?
    \item \textbf{RQ2:} Are memory-safety-oriented unit proofs effective at exposing memory-safety vulnerabilities?
    \item \textbf{RQ3:} How do \toolname's safety-oriented techniques contribute to its performance and cost?
    \item \textbf{RQ4:} Do \toolname-generated unit proofs differ from those of experts?
\end{itemize}

\subsection{Evaluation Setup}

\myparagraph{Subjects}
Following prior works on validating embedded software~\cite{amusuo_unit_2025-1}, we evaluate \toolname using four widely-used open-source embedded operating systems:
  \textit{Zephyr RTOS}, \textit{RIOT OS}, \textit{Contiki-NG} and \textit{FreeRTOS}.
These operating systems are substantial (\cref{tab:evaluation-subjects}), including task management, IPC, networking, and storage subsystems.
We considered including downstream embedded applications but their source code is often not publicly available~\cite{ritvik2025lemix}.

\myparagraph{Vulnerability selection}
\label{par:eval-set-vuln-selection}
To assess \toolname's effectiveness to expose security vulnerabilities, we identify and recreate 60 known CVEs in selected subjects.
To ensure reliable recreation, we extract CVEs affecting each software from the National Vulnerability Database and identify the first 20 CVEs where the vulnerable function still exists and the security advisory provided details of the vulnerability sink and the fixing commit. 
The FreeRTOS CVEs in our dataset did not provide the fixing commits and were excluded.
We re-expose each CVE by applying a patch to reverse the changes in the identified fixing commit.
For each CVE, we also record the vulnerability type, affected function, and sink location, to enable exposure detection.

\myparagraph{Verification targets}
Component-level verification require entry points through which the components would be verified.
We identify these entry points in three steps.
First, we include the 57 functions affected by the 60 selected CVEs~\footnote{3 functions contained 2 CVEs each}.
Second, to assess generalizability beyond known vulnerable functions, we randomly select an additional 100 functions across the four embedded OSes.
First, we identify five modules in each operating system that handle untrusted data, such as bluetooth, network and USB modules.
For each module, we perform an attack surface analysis to identify sources of untrusted data and the functions that process them.
We select 5 functions per module, yielding an additional 25 functions per OS.
Finally, we select 20 FreeRTOS functions with expert-written harnesses to enable head-to-head comparison in RQ4.
Our component entry point set thus has a total of 177 entry points.
\Cref{tab:evaluation-subjects} shows the parent modules containing the selected entry points. 

\myparagraph{Baselines.}
We compare \toolname against two categories of baselines.
First, we compare against alternative methods for producing unit proofs, including expert-written FreeRTOS proofs and proofs generated by vanilla coding agents (GPT Codex).
This comparison evaluates differences in verification choices and resulting outcomes.
Second, we compare against memory-safety verification techniques, using Seeker~\cite{takhar_memory-safety_2025}, a recent open-source verifier for memory safety of open programs.
Seeker is the closest prior system to our setting.
We defer conceptual comparisons with other verification methods to \cref{sec:related-work}.

\myparagraph{Hardware and AI models}
We run experiments on dedicated servers (32 virtual CPUs, 188GB of RAM).
We use gpt-5.3-codex through their API for our evaluation, as it ranked among the best AI models for coding~\cite{noauthor_swe-bench_nodate} at time of study.
We configure with the default temperature (1.0) and reasoning effort (high).
To assess generalizability to open-source models, we compare against Minimax M2.5 and GLM-5, two leading open-source models for coding~\cite{noauthor_swe-bench_nodate}.

\myparagraph{\toolname configuration}
We evaluate two \toolname configurations that differ in the maximum scope level used by resource-aware scope widening (\cref{subsec:resource-aware-scope-widening}).
Scope-1 (\textit{S1}) widens only to scope level of one (parent file containing entry point).
Scope-2 (\textit{S2}) uses depth two (parent file and all adjacent files).
For each verification run, we use a 30-minute timeout, a practical budget under which most component-level verification tasks complete.
We use the default configurations for the baseline techniques.


\subsection{RQ1: Are \toolname's Results Useful?}
\label{sec:rq1}

\cref{subsec:ps-success-criteria} identifies five properties that unit proofs must satisfy to support useful verification guarantees: structural validity, conclusiveness, verification coverage, result validity and maintainability.
RQ1 evaluates the first three and last properties.
We defer result validity to RQ2.

\subsubsection{Method}

We evaluate RQ1 on the full set of selected component entry points and use both \toolname configurations.
We run AutoSOUP-S1 on all verification targets and AutoSOUP-S2 only on targets with recreated CVEs because its wider scope increases generation and verification cost.
We terminate generation runs that exceed 24 hours.

For each generated unit proof, we measure whether generation completes, whether the proof compiles, whether it is semantically valid, and how long verification takes.
We then measure the verification outcome: the total lines of code in functions statically reachable from the unit proof, the proportion of those lines covered under the proof bounds and assumptions, the total number of instrumented memory-safety properties, and the proportion of those properties verified.

We also measure the cost and size of each unit proof.
For cost, we record total generation time and API cost.
For size, we record proof size in lines of code, including both harness function and all function models.
Following prior work on unit testing~\cite{robinson_scaling_2011, daka_modeling_2015}, we use unit proof size as a proxy for maintainability.

For comparison, we run Codex and Seeker on the verification targets with recreated vulnerabilities and report the corresponding measurements when available.
For Codex, we use a prompt that states the goal of unit proof creation and the success criteria in \cref{subsec:ps-success-criteria}, but gives no specific guidance on how to derive verification choices.
This baseline tests whether a general-purpose AI coding agent can independently produce unit proofs that support useful memory-safety guarantees.
For Seeker, we extract the compilation configurations from \toolname-generated unit proofs and the scripts provided in the Seeker artifact to compile and verify the target file.
To validate our setup, we sampled benchmarks from the Seeker codebase and confirmed that our instrumented programs matched the artifact versions and produced identical verification results.

\subsubsection{Result}

\myinlineparagraph{Unit proof validity:}
\cref{tab:rq1-harness-utility} compares the validity, verification outcomes, and generation cost of unit proofs produced by \toolname and CodexUP.
\toolname produced substantially more valid unit proofs than Codex: 93\%, 89.5\% and 91.7\% for scope levels 1 and 2 and the randomly selected targets, respectively, compared with 31.6\% for CodexUP.
In Seeker, of the 57 targets, 12 (21.1\%) returned an error while 23 (40.4\%) timed out.

Most Codex proofs were structurally invalid because Codex often failed to compile or verify the target, due to missing compilation, incomplete environment models or initially large loop bounds, and instead created simpler copies to verify.
Seeker more often returned errors when it could not process the source file and timed out more often because it assigned a uniform bound of 20 to all loops and did not create valid environment models that reduce the state space.
This shows that \toolname's structured refinement and validation are necessary to produce unit proofs that compile, reach the target, and complete verification.

\begin{table}
    \centering
    \caption{
    Comparison of unit proofs produced by \toolname at scope level 1 and 2 and by CodexUP. Random represents targets selected to assess generalizability. We report results from Seeker baseline in the prose.
    Below the double-line, we consider only the unit proofs from successful runs for each method.
    }

\begin{tabular}{lccc|c}
\toprule
\textbf{Metric}  & \textbf{S1} & \textbf{S2} & \textbf{Codex} & \textbf{Random}\\
\midrule
Num Targets  & 57 & 57 & 57 & 84~\footnote{16 test cases were disrupted due to campus-wide power outage during our experiments and have been excluded from our results} \\
Struct. valid (\%) & 93\% & 89.5\% & 31.6\% & 91.7\% \\
Verification completes (\%) & 96.5\% & 91.2\% & 70.2\% & 76.2\% \\
Generation succeeds (\%) & 93.0\% & 78.9\% & 98.2\% & 75\% \\
\midrule
\midrule
Verification time (s) & 137.9 & 249.1 & 37.4 & 17.7 \\
Avg compon. size (loc) & 126.1 & 418.9 & 132.5 & 107.7 \\
Avg covered size (loc) & 108.8 & 147.4 & 57.8 & 88.8 \\
\midrule
Avg num. properties (\#) & 382.4 & 1187.2 & 468.3 & 243.2 \\
Avg prop. verified (\#) & 369.5 & 1185.3 & 465.1 & 242.7 \\
Avg reported errors (\#) & 4.7 & 3.2 & 1.6 & 1.2 \\
\midrule
Avg gen. time (min) & 129.8 & 363.7 & 10.1 & 45.0 \\
Avg API cost (\$) & 3.0 & 5.9 & 0.8 & 1.8 \\
\midrule
Avg proof size (loc) & 34.1 & 40.1 & 50.7 & 27.3 \\
\bottomrule
\end{tabular}
    \label{tab:rq1-harness-utility}
\end{table}

The lower success rate of AutoSOUP-S2 relative to AutoSOUP-S1 is expected.
Increasing the scope level adds more adjacent code to the verification target, which increases program size, state space, and verification time.

\myinlineparagraph{Verification outcomes and assurance:}
Among valid unit proofs, \toolname, at scope levels 1 and 2, covered and verified 88.2\% and 155\%  more lines of code compared to Codex' 57.8 lines of code.
Data from Seeker is excluded as it does not provide coverage report.
\cref{tab:rq1-harness-utility} also provides the number of total and verified properties as reported by the tool. 
However, because it counts the properties in non-covered code as verified because no violation was produced, Codex substantially lower coverage led to a higher number of reported verified properties.
These results show that \toolname also outperforms frontier coding agents in generating unit proofs that achieves better verification coverage and provides stronger memory-safety assurances.

The remaining uncovered code was primarily caused by statements that were not statically reachable from the generated unit proof, even when their enclosing functions were reachable.
This effect becomes more pronounced as scope level increases and the proof includes functions from adjacent files.
\toolname nevertheless achieved higher coverage than CodexUP because its generated proofs are designed to admit all valid reachable states.
By contrast, CodexUP often introduced restrictive assumptions in its unit proofs that excluded feasible states, reducing both code coverage and the corresponding number of properties checked.

\myinlineparagraph{Unit proof generation cost:}
Generating \toolname unit proofs required 2.16 and 6.06 hours on average for scope levels 1 and 2, respectively, and cost \$3 and \$5.9 in API usage.
These costs are higher than CodexUP, but they produce substantially more valid and useful proofs.
They are also modest relative to prior human-driven unit proofing effort, where verifying 1,500 lines of code required about one person-month of work~\cite{chong_code-level_2020}.

\begin{figure}
    \centering
    \includegraphics[width=\linewidth]{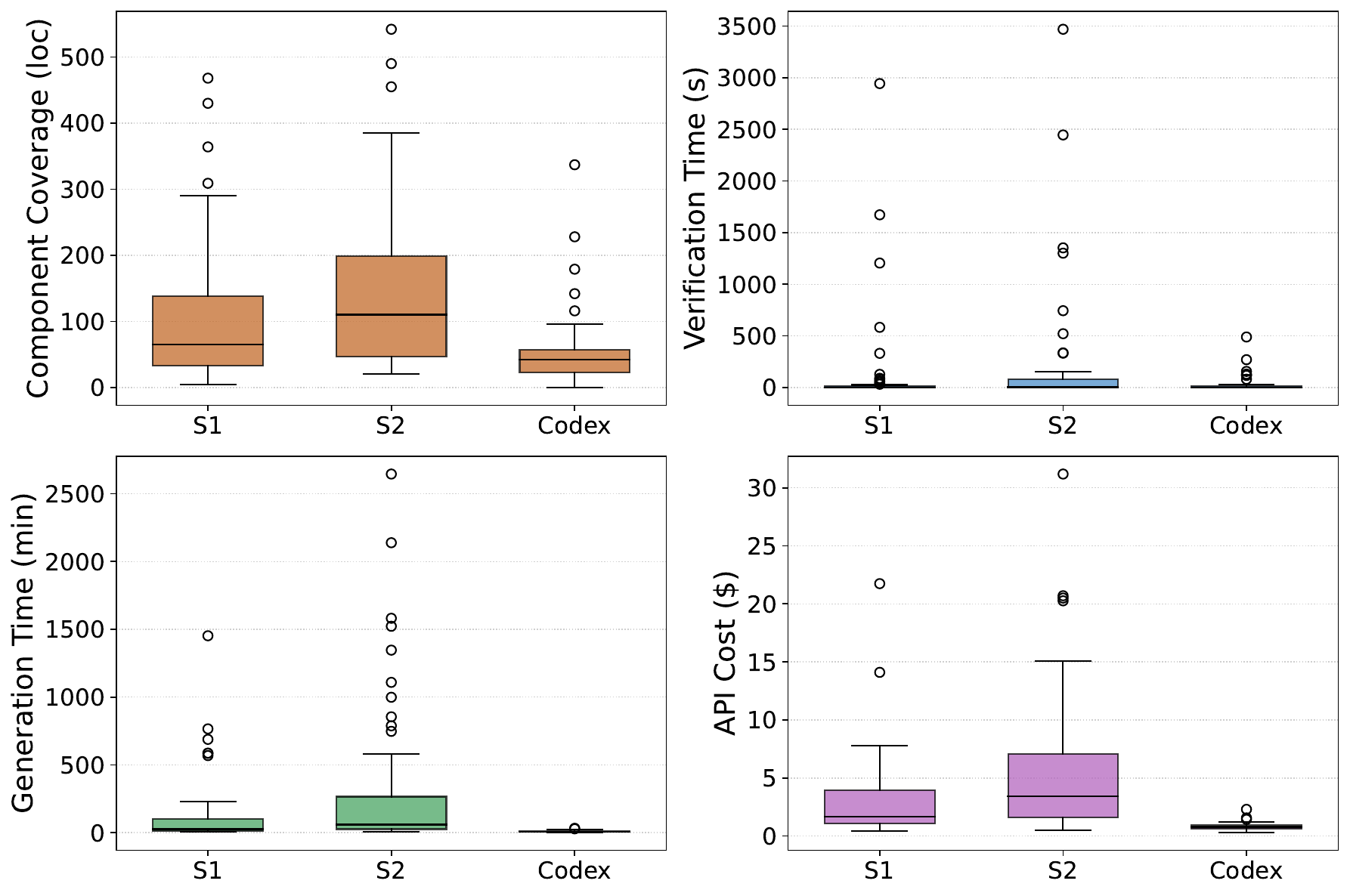}
    \caption{Distribution of program reachable LOC, verification time, development time, and API cost for the RQ1 unit proofs.}
    \label{fig:rq1-coverage}
\end{figure}

\cref{fig:rq1-coverage} shows that the cost of \toolname was distributed across the unit proofs.
Overall, increasing the scope level from 1 to 2 increased the size of verified code, together with the development costs.
Our data also showed these costs correlated closely with component size: larger component sizes took longer to verify, which increased the iterative-refinement-based development time.

\subsection{RQ2: Does \toolname Find Vulnerabilities?}
\label{sec:rq2}

\cref{subsec:ps-success-criteria} requires unit proofs to produce valid memory-safety results: their verification outcomes should reflect genuine memory-safety vulnerabilities in the verified component.
RQ2 evaluates whether unit proofs generated by \toolname expose known memory-safety vulnerabilities during verification.

\subsubsection{Method}

We evaluate RQ2 on verification targets with recreated vulnerabilities.
We compare \toolname against the baselines using the proportion of vulnerabilities exposed and the root causes of any missed vulnerabilities.

Counting a vulnerability as exposed is non-trivial.
During context-aware environment refinement (\cref{subsec:context-aware-env-refinement}), \toolname infers underapproximate preconditions that suppress reported property violations.
A precondition inferred for one violation may also suppress other violations that share the same root cause.
As a result, the violation location recorded by \toolname may differ from the CVE sink location defined in \cref{par:eval-set-vuln-selection}, even when both correspond to the same vulnerability.

We therefore count a vulnerability as \emph{exposed} if any of the following holds:
(i) \toolname reports a memory-safety error at the recorded CVE sink, and the inferred precondition can be violated at the component callsite;
(ii) \toolname reports a memory-safety error at a different sink, the inferred precondition can be violated at the callsite, and violating it would also trigger a property violation at the recorded CVE sink; or
(iii) \toolname reports a memory-safety error at the recorded CVE sink without inferring a precondition that suppresses the error.

We identify the first and third cases automatically by comparing recorded CVE sink locations with the memory-safety errors reported by \toolname.
We identify the second case manually by checking whether removing the inferred precondition produces a property violation at the recorded CVE sink.
Finally, we compare the proportion of CVEs exposed by \toolname, Codex, and Seeker, and manually categorize missed vulnerabilities to understand limitations of our techniques and those of the underlying bounded model checker.

\subsubsection{Result}

\myinlineparagraph{Exposure of known CVEs:}
\cref{tab:rq2-vuln-exposure} reports the proportion of recreated CVEs exposed by each method and summarizes the reasons for non-exposure.
AutoSOUP-S1 and AutoSOUP-S2 exposed 66.7\% and 65\% of CVEs, respectively.
In contrast, Codex and Seeker exposed only 28.3\% and 41.7\%~\footnote{For Seeker, due to limited information it provides, we counted any violation corresponding to the CVE sink line as exposure}.
These results show that \toolname exposes substantially more known vulnerabilities than both a general-purpose coding agent and the prior verification baseline.

\begin{table}
    \centering
    \caption{
    Exposure rate of known vulnerabilities.
    S1 and S2 represents \toolname configured at scope levels one and two respectively.
    }
    \begin{tabular}{lcccc}
    \toprule
        Metric & S1 & S2 & Codex & Seeker \\
        \midrule
        Number of CVEs & 60 & 60 & 60 & 60 \\
        Number exposed & 40 & 39 & 17 & 25 \\
        Unexploitable & 6 & 5 & 1 & - \\
        \midrule
        Compilation error & 3 & 4 & 6 & - \\ 
        Structural invalidity & 0 & 0 & 2 & - \\
        Resource exhaustion & 1 & 2 & 17 & - \\
        \midrule
        Limited scope & 3 & 0 & 2 & -\\
        Limited loop unwinding & 1 & 3 & 0 & -\\
        Inaccurate env. model & 2 & 0 & 14 & - \\
        \midrule
        Unsupported sink & 2 & 2 & 1 & - \\
        BMC field granularity & 2 & 4 & 0 & - \\
        \bottomrule
    \end{tabular}
    \label{tab:rq2-vuln-exposure}
\end{table}

\myinlineparagraph{Root causes of non-exposure:}
We classify non-exposure into three categories.
First, some unit proofs were invalid or inconclusive, preventing verification from producing a useful result.
Second, some proofs used insufficient verification choices, such as scopes that excluded vulnerability-relevant paths, loop bounds that were too small, or environment assumptions that ruled out vulnerable states.
Third, some cases were limited by the underlying bounded model checker, even when the relevant code and triggering conditions were present.

For AutoSOUP-S1, missed CVEs were primarily caused by limited scope, CBMC limitations, or vulnerabilities that were not triggerable from the project context.
The scope-related misses were resolved by AutoSOUP-S2, showing that scope widening is important for exposing vulnerabilities whose triggers cross file boundaries.

The CBMC limitations appeared in two forms.
First, CBMC reports a spatial memory-safety violation only when an access leaves the allocated object.
However, for structs, \toolname's property-guided loop-bound analysis computes the bound needed to overflow the destination field, not necessarily the containing object.
Thus, when a write overflows a struct field but remains within the same allocated struct object, CBMC does not report a violation.
Second, 2 vulnerabilities was missed because they required semantics that CBMC did not model. One was a memory-leak, missed because the corresponding custom deallocator is not supported. The second required timer semantics for exposure.

In cases we tagged unexploitable, \toolname reached the CVE sink and inferred the relevant precondition, but the precondition was valid in the component's calling context, either due to upstream constraints or default program configurations.

Codex missed CVEs mainly because its unit proofs were invalid, inconclusive, or over-constrained.
Its ad hoc proof generation often produced scopes that did not support conclusive verification or assumptions that excluded vulnerable states.
We could not investigate Seeker misses because no coverage report or unit proofs were produced.

\myinlineparagraph{Exposure of new vulnerabilities:}
During unit proof generation, \toolname scope levels 1 and 2 reported 63 and 99 potential memory-safety violations where inferred preconditions were violated.
After reviewing a subset of these reports, we confirmed 20 new externally triggerable vulnerabilities (\cref{tab:autoup-new-vulns}).
Nine of these are out-of-bound write vulnerabilities, with potentials to cause denial of service or arbitrary code execution.
Another 7 are out-of-bound reads, which can potentially lead to information disclosure attacks.
Details of one vulnerability is in \cref{sec:new-vulns}.
We reported all confirmed vulnerabilities to the corresponding maintainers.
One has been fixed and assigned a CVE.
Another 4 were fixed without CVE assignment because maintainers considered exploitation to require prior compromise of the downstream embedded application, which was outside their threat model.
The remaining reports are still under investigation.

\myinlineparagraph{Sample out-of-bound write vulnerability in RIOT-OS:}
\Cref{listing:new-vuln} illustrates a vulnerability in RIOT-OS \texttt{nanocoap.c} discovered by \toolname.
\toolname first generated a harness for the root function \texttt{coap\_opt\_put\_uri\_pathquery}, initializing \texttt{buf} and \texttt{string} as nondeterministic pointers with unconstrained sizes.
Stage~2 refined the proof bounds and models until verification reached the vulnerable \texttt{memcpy} on Line 33, where a memcpy write violation was exposed.
Finally, Stage~3 inferred a precondition on the input-buffer length that would eliminate the violation, but the validator traced the constrained string back to the nanocoap\_sock\_post public API and found no corresponding length check.
\toolname therefore classified the precondition as violable in the real environment and reported the \texttt{memcpy} out-of-bounds write vulnerability.

\begin{table}
    \centering
    \caption{New vulnerabilities discovered using \toolname.
    Each count is reported as \textit{Total (Contiki-NG, Zephyr, RIOT)}.}
    \begin{tabular}{lc}
    \toprule
    \textbf{Vulnerability Type} & \textbf{Count} \\
    \midrule
        Out-of-bound write & 9 (5, 1, 2) \\
        Out-of-bound read & 7 (2, 3, 2) \\
        Undefined shift behavior & 2 (1, 1, 0) \\
        Null pointer dereference & 1 (1, 0, 2) \\
        \midrule
        \textit{Total} & 20 (9, 5, 6) \\
        \bottomrule
    \end{tabular}
    \label{tab:autoup-new-vulns}
\end{table}

\subsection{RQ3: Ablation of \toolname's Techniques?}
\label{sec:rq3}

We ablate the techniques used to derive our safety-oriented unit proofs:
  verification scope, loop bounds, and environment models.

\subsubsection{Method}

During the RQ1 runs, we save a unit proof snapshot after each technique completes.
For each snapshot, we also record the technique's execution time and API cost.

We use these snapshots to measure how each technique changes the unit proof.
Specifically, we measure changes in verification scope, loop bounds, and environment models.
We then execute each snapshot and measure the resulting verification coverage, the number of reachable memory-safety properties, and the proportion of those properties verified.
This allows us to isolate how each verification choice affects verification outcomes and to estimate the marginal cost of each technique.

Finally, we compare generalizability to open-source models.
We execute the 37 targets in RIOT and Contiki-NG using \toolname configured at scope level 1 and equipped with the selected open-source models, Minimax M2.5 and GLM-5.
Similar to RQ1, we measure the validity of produced unit proofs, their verification outcomes and the generation cost.
\subsubsection{Result}

\myinlineparagraph{Contribution and cost of each technique:}
\cref{rq3:tab-contributions} summarizes how each \toolname stage changes the unit proof, affects verification outcomes, and contributes to generation cost.
Overall, the results show that the three techniques are complementary: each stage introduces a distinct class of verification choices and improves a different aspect of the final proof.

\begin{table}
    \centering
    \caption{Stage-wise contribution of \toolname's scope, loop-bound, and environment-modeling stages to harness structure, verification behavior, development time, and API cost.}

\begin{tabular}{lccc}
\toprule
\textbf{Metric} & \textbf{Stage 1} & \textbf{Stage 2} & \textbf{Stage 3} \\
\midrule
Harness Size (LOC) & 6.613 & 7.795 & 21.64 \\
Proof Size (LOC) & 18.74 & 25.76 & 44.81 \\
\midrule
\# Functions In Scope & 17.78 & 18.84 & 17.34 \\
\# Custom Loop Bounds & 0.004 & 0.909 & 0.842 \\
\# Variable Models & 0.044 & 0.068 & 3.424 \\
\# Function Models & 1.351 & 1.938 & 2.119 \\
Avg Function Model Size (LOC) & 2.954 & 3.602 & 5.402 \\
\midrule
Component Size (LOC) & 209.8 & 220.8 & 196.3 \\
Verification Coverage (LOC) & 94.06 & 139.6 & 115.8 \\
Total Properties & 542.8 & 580.0 & 539.0 \\
Verified Properties & 412.3 & 412.8 & 531.6 \\
\midrule
Generation Time (min) & 6.054 & 39.99 & 141.5 \\
API Cost (\$) & 0.296 & 0.743 & 2.602 \\
\bottomrule
\end{tabular}

    \label{rq3:tab-contributions}
\end{table}

Stage 1 (resource-aware scope widening) determines the reachable functions included in the proof and introduces models for undefined functions.
These choices remain largely unchanged in later stages.
This stage is also the cheapest because it is implemented primarily using deterministic program analysis.

Stage 2 (property-guided loop refinement) increases the number of per-loop bounds.
This expands the explored behavior of the component, which increases both verification coverage and the number of reachable memory-safety properties.
At this stage, up to 71.04\% of reachable properties on average are violated under the unconstrained environment, showing that loop unwinding exposes safety-relevant behaviors that must later be checked or constrained.
This stage takes 39.99 minutes and costs \$0.743 on average.

Stage 3 (context-aware environment refinement) primarily refines the variable and function models to eliminate infeasible executions while preserving safety-relevant behavior.
This allows 98.63\% of reachable properties to be verified and produces the environment assumptions under which the verification holds.
It is also the most expensive because \toolname infers and validate preconditions for reported property violations one by one.

Together, these results explain \toolname's performance.
Scope widening determines what code is analyzed, loop unwinding exposes safety-relevant behaviors within that code, and environment refinement separates feasible violations from behaviors ruled out by the component context.

\myinlineparagraph{Generalizability to open-source models:}
\cref{tab:rq3-model-comparison} in \cref{sec:appendix:AdditionalData} evaluates \toolname with open-source models.
Across 37 test cases, unit proof generation succeeded for 78.4\% with GLM-5 and 48.6\% with Minimax M2.5, compared to 93\% with GPT-5.3-Codex.
Their average costs were \$4.9 and \$0.7, respectively, compared to \$3 for GPT-5.3-Codex.
These results show that \toolname can generalize to open-source models, but its performance depends on model capability.
As open-source models improve, we expect corresponding gains in \toolname's performance with them.

\subsection{RQ4: \toolname vs. Expert-Written Proofs?}
\label{sec:rq4}

RQ4 compares unit proofs generated by \toolname with existing FreeRTOS proofs.
We restrict this analysis to FreeRTOS because it is the only evaluation subject with unit proofs developed by project maintainers.

\subsubsection{Method}

We first compare verification choices and outcomes quantitatively.
For verification choices, we measure proof size, verification-scope size, distinct loop bounds, and the number of variable and function models.
For outcomes, we measure verification time, component size, verification coverage, the number of verified properties, and the number of violated properties.

We then qualitatively analyze the environment models using the taxonomy from prior work~\cite{amusuo_unit_2025-1}.
Variable models fall into four categories: null-pointer preconditions (\texttt{p != NULL}), pointer-offset preconditions (\texttt{p2 = p1 + offset}), variable-constant preconditions (\texttt{var >= CONSTANT}), and variable-variable preconditions (\texttt{var1 >= var2}).
Function models fall into three categories: Type~1 models with no preconditions, Type~2 models with preconditions only on return values, and Type~3 models with preconditions on inputs or global variables.
We classify each model and compare the assumptions encoded by \toolname-generated and expert-written proofs.

Finally, we conduct a case study of three randomly selected functions with varying proof sizes.
We inspect their loop bounds and environment models to explain observed differences and their implications for verification.

\subsubsection{Result}
\begin{figure}
    \centering
    \includegraphics[width=\columnwidth]{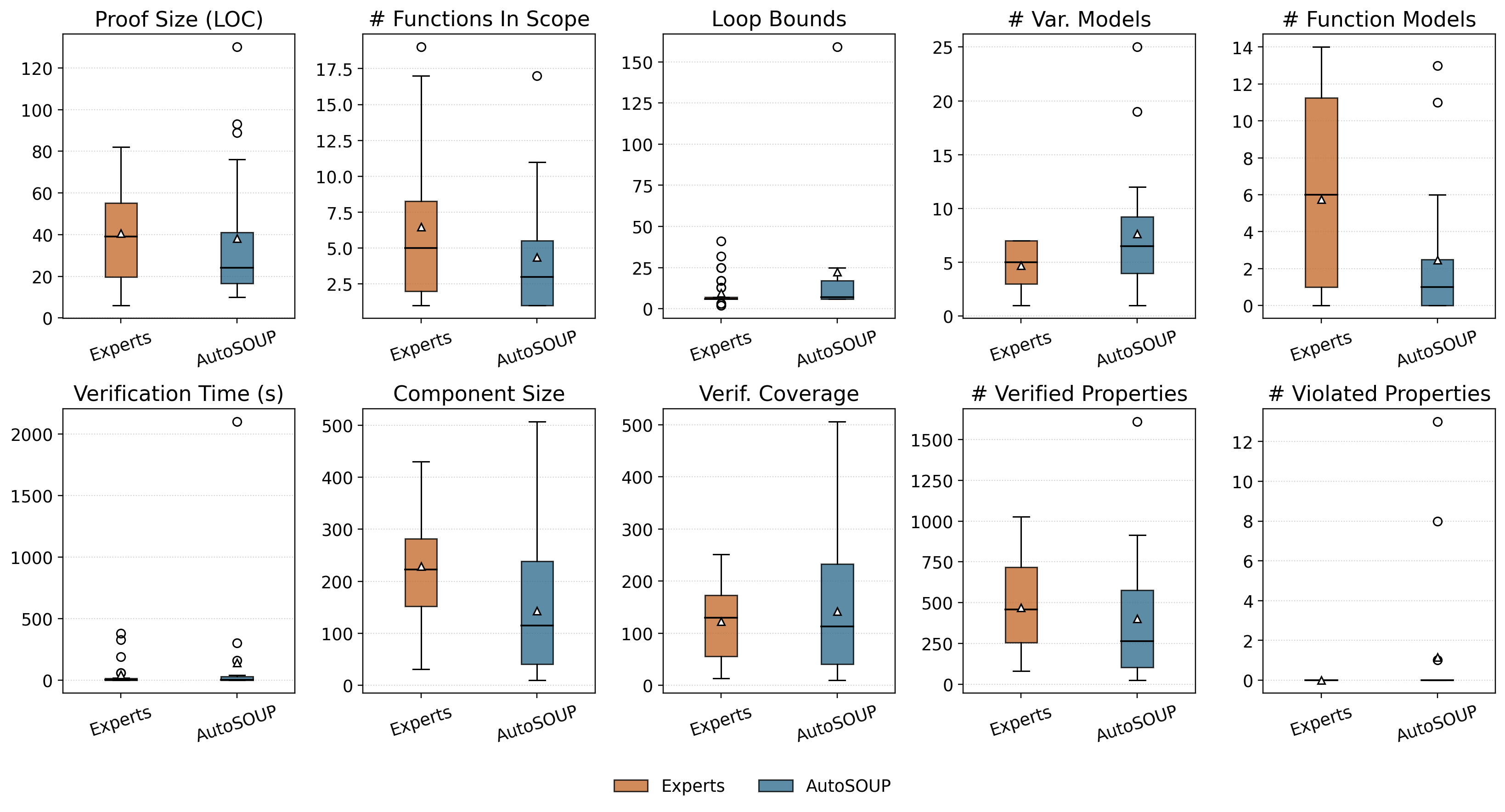}
    \caption{Comparing \toolname's unit proofs with expert-written ones. Top measures verification choices. Bottom measures verification outcomes.}
    \label{fig:rq4-quatitative-comparison}
\end{figure}

\myinlineparagraph{Quantitative comparisons:}
\cref{fig:rq4-quatitative-comparison} compares the verification choices and outcomes.
Overall, \toolname-generated unit proofs are smaller, use smaller verification scopes, include more variable models and fewer function models, and achieve comparable verification coverage.
This difference stems from two design choices.
First, \toolname uses preconditions to constrain environment variables, whereas FreeRTOS proofs often call project functions to initialize or constrain state, such as the proof for \texttt{vDHCPProcess} using \texttt{prvCreateDHCPSocket} to initialize DHCP sockets.
Second, FreeRTOS proofs often replace same-file callees with function models and reuse shared function models across proofs.
In contrast, \toolname keeps same-file callees in scope and creates models only for functions outside the entry point's parent file, resulting in fewer function models.

\myinlineparagraph{Qualitative comparisons:}
\cref{tab:rq4-model-categories} shows the distribution of model categories.
Although \toolname introduces 90\% more preconditions than the expert-written proofs, most are simple: non-null pointer constraints or fixed upper and lower bounds on environment variables needed for memory safety.
Thus, \toolname-generated environment models are comparable to expert-written models, making them auditable and maintainable.
Case studies comparing specific unit proofs are in \cref{sec:appendix:case-studies}. Sample \toolname-generation unit proof is in \cref{listing:prvcheckoptions-harness}.

\begin{table}
    \centering
    \caption{Categorization of environment models from unit proofs. Data is aggregated from 20 unit proofs each. Top section represent variable model categories, while bottom section is for function models.}
    \begin{tabular}{ccc}
    \toprule
        \textbf{Model category} & \textbf{Expert-written} & \textbf{\toolname} \\
        \midrule
        pointer-not-null & 65 & 62 \\
        pointer-offset & 2 & 1\\
        variable-constant & 38 & 70\\
        variable-variable & 3 & 8 \\
        \midrule
        Type~1 models & 58 & 23 \\
        Type~2 models & 18 & 19 \\
        Type~3 models & 0 & 2 \\
        \bottomrule
    \end{tabular}
    \label{tab:rq4-model-categories}
\end{table}

\section{Discussion and Related Work}
\label{sec:Discussion}

We discuss implications for practitioners and researchers and compare our work to related lines of research.

\subsection{Discussion}

\myparagraph{Guarantees, cost, and practicality.}
\toolname occupies a practical middle ground in the software-verification landscape.
Deductive verification~\cite{filliatre_deductive_2011} can prove rich functional and memory-safety properties, but requires substantial specifications, annotations, and proof effort.
Expert-written unit proofs for bounded model checking~\cite{chong_code-level_2020, wu_verifying_2024} can provide useful memory-safety guarantees, but require experts to define scopes, loop bounds, inputs, and environment models.
Automated analyzers such as Infer~\cite{calcagno_infer_2011} operate at lower cost, but target narrower classes of properties than the bounded memory-safety guarantees studied in this paper.
In contrast, \toolname automates these choices and makes them explicit, while providing bounded memory-safety guarantees.
RQ2 (\cref{sec:rq2}) shows that this tradeoff is useful in practice: \toolname exposes 66.7\% of recreated CVEs and reports the assumptions under which each verified component is memory safe.
It is also substantially cheaper than prior industry experience where expert-written proofs take one engineer-month to cover 1500 lines of code.
Although \toolname does not replace stronger verification methods for safety-critical certification, it lowers the cost of \msv for broader software teams.

\myparagraph{Integration in developer workflow.}
\toolname makes component-level verification more practical for low-level software projects.
RQ1 (\cref{sec:rq1}) shows that it can generate unit proofs and verify components in realistic embedded software.
This supports a practical \textit{shift-security-left} workflow~\cite{pitchford_shift_2021}: developers can obtain formal evidence about component memory safety during development, when defects are cheaper to diagnose and fix~\cite{noauthor_software_defect_nodate}.
Our aeronautics industry partners noted that this capability is especially valuable in long development cycles, where products may take up to seven years to complete.
In addition, by making environment assumptions explicit in the unit proof, \toolname also exposes the interface contracts under which a component is memory safe.
This will help developers identify and validate the caller-side obligations whose violation often leads to component-interface vulnerabilities~\cite{lefeuvreAssessingImpactInterface2023}.

\myparagraph{Implications and future directions.}
\toolname also illustrates a path for trustworthy AI-assisted software engineering.
As developers increasingly use AI agents to generate and modify code, tools like \toolname can provide machine-checkable evidence that generated components are memory safe under explicit assumptions.
Its \textit{LLM-as-function-call} architecture also allows model improvements to strengthen LLM-driven subtasks without weakening the deterministic orchestration and validation framework.
This matters because formal guarantees depend on the assumptions encoded in the verification task~\cite{fonseca_empirical_2017}, and automatically judging the correctness of these assumptions remains an open problem~\cite{amusuo_unit_2025}.

Looking forward, extending \toolname from component-level memory-safety verification to broader project-level assurance requires three advances.
First, loop unwinding should be combined or replaced with loop invariant generation~\cite{pirzada_llm-generated_2024} to expose vulnerabilities whose required bounds exceed the resource budget.
Second, whole-project verification requires techniques that select effective component entry points, reduce bound or environment refinement cost with program analysis, and compose component-level results into project-level guarantees~\cite{cho_blitz_2013, bensalem_compositional_2010}.
Third, generalizing beyond memory safety requires techniques that automatically infer system and security properties that can be verified using bounded model checking. Novel AI-driven techniques can help infer them from existing specifications, comments, and unit tests~\cite{le-cong_can_2025}, then encode them as safety properties for verification.

\subsection{Related Work}
\label{sec:related-work}

We situate our work within three lines of research.

\myparagraph{(1) Automating component-level BMC}
Prior work has explored automating bounded model checking through compositional verification~\cite{cho_blitz_2013, ivancic_dc2_2011}.
These techniques decompose programs along call-graph or control-flow edges, infer preconditions under which each fragment is safe, and discharge the resulting obligations using assume-guarantee reasoning~\cite{cobleigh_learning_2003}.
However, it remains unclear whether these techniques scale to real-world systems with large components, diverse coding patterns, and complex inter-component relationships.
Other work targets open programs, where components have undefined dependencies, and refines their environments using expert-provided precondition templates~\cite{takhar_memory-safety_2025, das_angelic_2015}.
Our evaluation shows that such templates can be imprecise and incomplete, compromising the resulting verification guarantees.
In contrast, \toolname automates industry-adopted component-level verification workflow and derives the scope, loop bounds, and environment models needed for practical \msv.
The derived choices can also support compositional and open-program verification methods.

\myparagraph{(2) LLMs and the verification-oracle problem}
Large language models are increasingly used to automate software-verification tasks, including tactic generation for deductive verification~\cite{kasibatla_cobblestone_2024, yang_autoverus_2024, first_baldur_2023}, precondition and loop-invariant inference for bounded model checking~\cite{pirzada_llm-generated_2024}, and repair of verification failures~\cite{tihanyi_new_2025}.
In these settings, the LLM usually acts as the main driver and the task has a clear success oracle: the generated artifact is accepted if verification succeeds.
\toolname addresses a harder oracle problem because successful verification is not sufficient when unjustified choices of scope, bounds, or environment assumptions can make verification succeed without providing appropriate guarantees.
Thus, \toolname uses LLMs as assistants inside program-analysis-driven workflows: deterministic orchestrators issue small tasks aligned with specific algorithms and validate each result before incorporating it into the unit proof.
This architecture may also apply to other security and software-engineering tasks, such as fuzzing, where AI-generated harnesses can compile and achieve coverage yet still produce false crashes or invalid results~\cite{zhang_how_2024, amusuo_falsecrashreducer_2025, xu_ckgfuzzer_2025}.

\myparagraph{(3) Security vulnerability detection}
\toolname complements static analysis~\cite{chen_static_2015, distefano_scaling_2019} and dynamic analysis~\cite{amusuo_systematically_2023, yun_fuzzing_2023, zhang_firmware_2020} for vulnerability detection.
Unlike these techniques, \toolname produces auditable unit proofs that specify the scope, bounds, and assumptions under which a component is memory safe.
These proofs can be rechecked to obtain bounded formal evidence of memory safety.
Because \toolname operates statically on source code, it is also well suited to embedded software, where dynamic analysis is often limited by emulation, rehosting, and peripheral-modeling challenges~\cite{maier_basesafe_2020, zheng_efficient_2022, feng_p2im_2020, bley_protocol-aware_2025, fasano_sok_2021}.

\section{Conclusion}
\toolname makes component-level bounded model checking more practical by automatically constructing unit proofs that verify a component's memory safety, while documenting the bounds and environment assumptions for the guarantees. 
It combines LLMs with a deterministic, incremental workflow to balance utility with reliability.
Our evaluation on four substantial embedded RTOSes
suggest automated harness generation as an incremental path toward security guarantees for real-world embedded software.

\ifARXIV
\section*{Acknowledgments}
Davis and Machiry acknowledge funding from Rolls Royce.
Amusuo and Anandayuvaraj acknowledge funding from the Qualcomm Innovation Fellowship.
All authors acknowledge API credits from OpenAI.
\fi

\cleardoublepage
\appendix

\cleardoublepage
\bibliographystyle{plainurl}
\bibliography{bib/main,bib/software-validation,bib/software-verification,bib/software-security}

\clearpage

\section*{Outline of Appendices}

\noindent
The appendix contains the following material:

\begin{itemize}
\item \cref{app-sec:ethics} Ethics analysis
\item \cref{app-sec:open-science} Open science information
\item \cref{app-sec:extended-algorithms} Extended version of \toolname algorithms
\item \cref{sec:appendix:case-studies} Unit Proof Case Studies
\item \cref{app-sec:memory-safety-properties} Instrumented memory safety properties

\item \cref{sec:new-vulns}: Sample vulnerability disclosed as a result of this work.
\item \cref{sec:appendix:AdditionalData}: Additional experimental results.



\end{itemize}

\section{Ethical Considerations}
\label{app-sec:ethics}
This section describes the ethical considerations of our work.
\toolname's primary ethical consideration is that it can be used to identify defects in a software system, resulting in a standard ``dual use'' scenario posing both harms and benefits.
We followed the guidance of Davis \etal in conducting our stakeholder-based ethics analysis~\cite{davis2025guide}.

\subsection*{Stakeholders}

\myparagraph{Direct stakeholders}
\begin{itemize}
    \item \textbf{Software maintainers and developers.}
    Engineers who use \toolname to analyze their own codebases for vulnerabilities and correctness issues.
    They directly interact with the tool and may act on its findings.
    \item \textbf{System operators and organizations.}
    Teams responsible for deploying and operating software systems evaluated using \toolname, including embedded, infrastructure, or safety-critical systems.
    \item \textbf{The research and development team.}
    Authors and maintainers of \toolname, who may face legal, professional, or reputational risks related to vulnerability discovery, disclosure practices, or downstream misuse.
    \item \textbf{Adversaries.}
    Malicious actors who could use \toolname, or techniques disclosed in the paper, to systematically discover vulnerabilities in target software.
\end{itemize}

\myparagraph{Indirect stakeholders}
\begin{itemize}
    \item \textbf{End users of affected software.}
    Individuals or organizations who rely on software analyzed using \toolname and may be impacted by vulnerabilities or by mitigations applied as a result of \toolname’s use.
    \item \textbf{Downstream software ecosystem.}
    Maintainers and users of libraries, dependencies, or products that incorporate code analyzed or modified using \toolname.
    \item \textbf{Vulnerable populations.}
    Groups disproportionately harmed by software exploitation, such as users of safety-critical, medical, industrial, or civic infrastructure.
    \item \textbf{Broader public.}
    Society at large, insofar as widespread exploitation or mitigation of vulnerabilities affects trust in software systems and digital infrastructure.
    \item \textbf{Research and security community.}
    Other researchers and practitioners who may reuse, extend, or operationalize \toolname’s techniques.
    \item \textbf{Policymakers and standards bodies.}
    Government and industry policymakers have asked for greater use of formal methods.
    Our work is intended to shift their cost/benefit calculus.
\end{itemize}

\subsection*{Potential Harms and Mitigating Factors}

\begin{itemize}
    \item \textbf{Exploitation risk.}
    \toolname could be used by adversaries to more efficiently identify exploitable vulnerabilities, including zero-day vulnerabilities, before patches are developed or deployed.
    \item \textbf{False positives.}
    When used by engineers, false positives in \toolname may waste their time. 
    Our evaluation characterized the false positives resulting from \toolname to let them make an informed adoption decision. 
    \item \textbf{Operational and economic impact.}
    Findings may require costly remediation, downtime, or architectural changes for system operators and organizations.
    \item \textbf{False confidence or misuse.}
    Over-reliance on \toolname’s outputs may lead developers to overlook other vulnerability classes or to misinterpret partial guarantees as comprehensive security.
    To reduce the risk of overgeneralization by engineers, our work characterizes the class of vulnerabilities exposed by \toolname, and the extent of false negatives.
    \item \textbf{Risk to researchers.}
    The research team may face legal or reputational consequences if \toolname is misused, misconfigured, or perceived as facilitating harm.
\end{itemize}

\subsection*{Potential Benefits}

\begin{itemize}
    \item \textbf{Improved software security.}
    \toolname enables engineers to detect and remediate vulnerabilities earlier and more systematically, reducing the likelihood of exploitation.
    \item \textbf{Support for higher-assurance engineering.}
    By producing structured analysis artifacts (\eg unit-level evidence or proofs), \toolname can assist organizations in meeting regulatory, safety, or compliance requirements.
    \item \textbf{Knowledge transfer and standardization.}
    The tool and associated research disseminate best practices for systematic vulnerability analysis and verification.
    They may change the standard cost/benefit analysis for the use of bounded model checking.
    \item \textbf{Research advancement.}
    \toolname contributes to the scientific understanding of systematic vulnerability detection and verification, enabling further defensive research.
\end{itemize}

\subsection*{Judgment}

In our judgment, the potential benefit to software cybersecurity outweighs the risks posed by our work.
We understood the ethical framing of our work at the outset of our study.
We did not observe any new concerns during the research conduct.
We therefore proceeded with submission to USENIX.


\section{Open Science}
\label{app-sec:open-science}
An anonymized artifact for our submission is available at:

https://anonymous.4open.science/r/AutoSOUP.

This artifact consists of:
\begin{enumerate}
\item The implementation of \toolname, consisting of both the conventional software and the prompts for the LLM components. We used an agentic framework that facilitates the use of open-source LLMs rather than commercial ones.
\item The organized benchmark of real-world CVE patches in open-source embedded operating systems, derived from the work of Amusuo \etal~\cite{amusuo_unit_2025-1}.
\item The evaluation automation that permit the replication of our results. 
\item The data associated with the evaluation automation that is used to create the tables and figures in this manuscript.
\item Links to the CVEs and defect repairs resulting from this research, as they emerge.
\end{enumerate}

In short, we provide everything we used to create our work and collect data.
We acknowledge that our use of OpenAI's frontier models complicates the replicability of this work.
We spent several thousand dollars to perform our experiments, and an independent replication would require a similar outlay of funds --- and the results will change because of the evolving nature of frontier models.

For more information, we refer the reviewers to the README of that artifact.

\section{Detailed AutoSOUP Algorithms}
\label{app-sec:extended-algorithms}

Here we provide detailed versions of the core \toolname algorithms.

\subsection{Resource-Aware Scope Widening}
\label{app:subsec:resource-aware-scope-widening}

Resource-aware scope widening derives the verification scope \(S_c\) by incrementally adding code that may contain checkable memory-safety properties and are semantically related while keeping verification within the resource budget \(R\).
Its goal is to increase the component behavior checked against \(Q\) without making verification intractable or the resulting unit proof difficult to audit.

Following prior work~\cite{takhar_memory-safety_2025}, we use source files as the unit of scope expansion.
This choice reflects the convention that related functions are often colocated in the same file.
File-level widening therefore preserves useful semantic context while keeping each expansion step coarse enough to manage.

The technique proceeds in three steps (\cref{alg:resource-aware-scope-widening}).

\begin{algorithm}

\caption{Resource-aware scope widening. 
The algorithm incrementally expands the verification scope at the file level and returns the largest scope whose provisional verification instance remains within the resource budget.}
\label{alg:resource-aware-scope-widening}
\small

\KwIn{Software system $S$, component entry point $C_e$, maximum scope level $d_{\max}$, resource budget $R$}
\KwOut{Verification scope $S_c$, loop bounds $B$, env. model $E$}

\Fn{\texttt{ResourceAwareScopeWidening}$(S, C_e, d_{\max}, R)$}{
    $S_c \leftarrow \texttt{AllFunctionsIn}(\texttt{FileOf}(C_e))$\;
    $B \leftarrow \texttt{InitBounds}(S_c, 1)$\;
    $E \leftarrow \texttt{InputModel}(C_e) \cup \texttt{ModelExternalCallees}(S_c)$\;

    \If{$\neg \texttt{WithinBudget}(S_c, B, E, R)$}{
        \KwRet $(\emptyset, \emptyset, \emptyset)$\;
    }

    \For{$d \leftarrow 1$ \KwTo $d_{\max}$}{
        $S_c' \leftarrow \texttt{WidenByOneFileLevel}(S_c)$\;
        $B' \leftarrow \texttt{InitBounds}(S_c', 1)$\;
        $E' \leftarrow \texttt{InputModel}(C_e) \cup \texttt{ModelExternalCallees}(S_c')$\;

        \If{$\neg \texttt{WithinBudget}(S_c', B', E', R)$}{
            \KwRet $(S_c, B, E)$\;
        }

        $(S_c, B, E) \leftarrow (S_c', B', E')$\;
    }

    \KwRet $(S_c, B, E)$\;
}

\BlankLine

\Fn{\texttt{WidenByOneFileLevel}$(S_c)$}{
    $F \leftarrow \texttt{FilesOf}(S_c)$\;
    $F_{\mathit{adj}} \leftarrow \texttt{FilesContainingCalleesFrom}(S_c)$\;
    \KwRet $\texttt{AllFunctionsIn}(F \cup F_{\mathit{adj}})$\;
}
\end{algorithm}

\myparagraph{Step 1: Initialize the scope, bounds, and input model}
We first add the file containing the component entry point \(C_e\) to \(S_c\), initialize all loop bounds in the current scope to~1, and construct an input model for \(C_e\).
The input model follows the entry-point type signature.
Primitive arguments receive unconstrained symbolic values over their full type range, while pointer arguments are initialized to valid allocated objects containing unconstrained values.
The resulting unit proof \(U(V)\) contains this input model and a call to \(C_e\).

We use an LLM agent to synthesize this input model and recover the configurations to compile the entry point's parent file.
Although the input model follows a fixed template, compiling it with the target file requires project-specific headers, include paths, macros and mandatory program configurations.
These requirements are difficult to recover reliably with fixed rules across diverse C projects~\cite{toman_taming_2017}.
Following the \textit{LLM-as-Function-Call} architecture, we mechanically validate the result: the proof must compile, call \(C_e\), and introduce no preconditions beyond the intended type-based initialization.

\myparagraph{Step 2: Model external calls}
We next identify call edges that cross the current scope boundary and replace their targets with simple type-based models.
From the compiled unit, we recover the call graph and identify edges to undefined callees and their return types.
We use an LLM agent to synthesize models for these undefined callees following strict guidelines, integrate them to the unit proof and ensure the resulting unit proof remains structurally valid.
Primitive returns are also modeled as unconstrained symbolic values, while pointer returns are modeled as valid allocated objects containing unconstrained values.

These unconstrained models preserve all possible states that may be returned by the excluded functions, including values that may violate \(Q\).
Validly-allocated pointer returns avoid irrelevant invalid-pointer states that cause state-space explosion and increase verification cost.
As in Step~1, an LLM agent generates and integrates the models, while deterministic checks ensure that the unit proof remains semantically valid.

\myparagraph{Step 3: Widen the scope}
After constructing the provisional instance \(V=(S_c,B,E)\), we check it against the configured resource budgets.
If verification remains within \(R\), we widen \(S_c\) by adding all reachable functions from files that contain definitions of previously excluded and modeled callees.
We use a pre-indexed database of \(S\) to locate candidate files.
When multiple files define functions with the same name and signature, we select the file closest to the in-scope caller by longest common path prefix.
The LLM agent is finally used to recover the compilation configuration for newly added files.

We repeat external-call modeling and scope widening until verification exceeds \(R\) or no additional files can be added.
We use three budgets: verification time, memory resources, and file depth.
The file depth budget bounds the number of files whose functions can be added to the verification scope so as to keep the unit proof generation cost reasonable.

\subsection{Property-Guided Loop Bound and Model Refinement}

A detailed description is in the main body of the manuscript.
We do not expand on it here.

\subsection{Context-Aware Environmental Model Refinement}
\label{app:subsec:model-refinement}

Property-guided refinement in \cref{subsec:property-guided-bound-refinement} maximizes exposure of violations of \(Q\) using unconstrained models \(E\).
However, violations found under unconstrained models may be infeasible in the broader system \(S\).
For example, the assertion on Program Line 15 in \cref{listing:sample-program-and-proof} will be violated by an input model that provides a null pointer, even though the actual caller only provides statically-defined arrays.
Context-aware environment refinement separates these infeasible property violations caused by overly permissive environment assumptions from genuine memory-safety errors.

This technique operates in two steps, illustrated in \cref{alg:context-aware-env-refinement}.

\setcounter{AlgoLine}{0}
\begin{algorithm}
\caption{Context-aware environment model refinement.
The algorithm infers underapproximate preconditions for violated memory-safety properties, validates them against calling contexts of the component entry point, and reports caller-feasible violations as memory-safety errors.}
\label{alg:context-aware-env-refinement-redux}
\small

\KwIn{Software system $S$, component entry point $C_e$, Environment model $E$, Property violations $Q_v$}
\KwOut{Refined environment model $E$, memory-safety error set $\mathcal{M}$}

\Fn{\texttt{ContextAwareEnvRefinement}$(S, C_e, E, Q_v)$}{
    $\mathcal{M} \leftarrow \emptyset$\;
    $\mathcal{W} \leftarrow \texttt{ParseViolationReport}(Q_v)$\;
    \tcp{$\mathcal{W}$ contains tuples $(q,w(q))$}

    \ForEach{$(q,w(q)) \in \mathcal{W}$}{
        $\phi \leftarrow \texttt{InferApproxPrecondition}(E,q,w(q))$\;
        $(\phi', \mathcal{B}) \leftarrow \texttt{ValidatePrecondition}(S, C_e, q, \phi)$\;

        $E \leftarrow E \cup \{\phi'\}$\;
        $\mathcal{M} \leftarrow \mathcal{M} \cup \mathcal{B}$\;
    }

    \KwRet $(E,\mathcal{M})$\;
}

\BlankLine

\Fn{\texttt{ValidatePrecondition}$(S, C_e, q, \phi)$}{
    $\mathcal{B} \leftarrow \emptyset$;
    $\phi' \leftarrow \phi$\;
    $\mathcal{C} \leftarrow \texttt{CallsitesOf}(C_e, S)$\;

    \ForEach{$c \in \mathcal{C}$}{
        $\psi_c \leftarrow \texttt{PathConstraints}(c, S)$\;

        \If{$\psi_c \not\models \phi'$}{
            $\hat{\phi} \leftarrow \texttt{WeakenPrecondition}(\phi',\psi_c)$\;

            \uIf{$\texttt{SatisfiesProperty}(\hat{\phi},q)$}{
                $\phi' \leftarrow \hat{\phi}$\;
            }
            \Else{
                $\mathcal{B} \leftarrow \mathcal{B} \cup \{(q,\psi_c,\phi')\}$\;
            }
        }
    }

    \KwRet $(\phi',\mathcal{B})$\;
}
\end{algorithm}

\myparagraph{Step 1: Infer underapproximate weakest preconditions}
Following prior counter-example-guided environment refinement approaches, we infer preconditions from counterexamples that suppress violations of memory-safety properties.
These preconditions need not be logically weakest, since weakest-precondition inference is often computationally expensive~\cite{cho_blitz_2013}.
Instead, they must be weak enough to preserve the target property while avoiding unnecessary restrictions on safe states.
For example, in \cref{listing:sample-program-and-proof}, the property on line~16 requires the loop index to remain within the size of \texttt{dst}.
The inferred precondition \(ret \le 10\) constrains the value returned by \texttt{get\_record\_count\_m2()} and therefore the number of loop iterations.
Further weakening this precondition by allowing larger values of \(ret\) violates the property.

We first parse the verification report to extract each violated property \(q \in Q\), its location \(\mathit{loc}(q)\), and its counterexample witnesses \(w(q)\).
For each tuple \((q, \mathit{loc}(q), w(q))\), we prompt an LLM agent to infer a precondition that, when added to \(E\), keeps \(\mathit{loc}(q)\) covered but suppresses the violation of \(q\).
The prompt guides the agent to identify the violated condition, propagate it backward through dataflow and path constraints, and stop at the external model or input responsible for the value.

In \cref{listing:sample-program-and-proof}, the agent first identifies the violated condition \(i < ObjectSize(dst)\), which simplifies to \(i < 10\).
It then propagates this condition through the loop condition on program line~13 to derive \(n \le 10\).
Finally, it propagates the constraint to the \texttt{get\_record\_count\_m2()} model and derives \(ret \le 10\).
The agent can inspect witnesses, navigate code, and test candidate preconditions.
We accept a candidate only if it suppresses the target violation without reducing structural validity, conclusiveness, coverage, or the number of checked properties.

\myparagraph{Step 2: Validate and refine against calling contexts}
The resulting precondition may overgeneralize and exclude valid caller states that would not violate \(q\).
As a result, we validate each accepted precondition against the calling contexts of \(C_e\) in \(S\).
Using a pre-indexed call graph, we identify callsites of \(C_e\) or the actual implementations of modeled functions.
For each callsite or implementation, we use an LLM agent to identify the constraints along execution paths that reach the callsite and check whether those constraints can violate the inferred precondition.

This validation produces three outcomes.
If a calling path violates the precondition but still satisfies \(q\), the agent weakens and revalidates the precondition.
If a calling path violates the precondition and triggers \(q\), we report the path as a feasible memory-safety error in \(S\).
If the precondition holds, we report the property as verified.
All preconditions are added to the unit proof's environment model \(E\).
They become explicit, auditable assumptions under which the verified component is memory safe.

\section{Unit Proof Case Studies}
\label{sec:appendix:case-studies}

\begin{listing}[t]
\centering
\caption{
    \toolname-generated unit proof for \texttt{prvCheckOptions}. The harness specifies the input model. It initializes the input socket and network buffer arguments. It also identifies conditions for memory safety: the ethernet frame \texttt{ethbuf\_len} is greater than the size of ethernet header (Line 24), the specified length value (Line 28) and the combined size of the ethernet, IP and TCP headers (Line 33 -- 41). The uxIPHeaderSizePacket function model specifies the memory-safety-relevant behavior of the function: that it returns the size of IP header corresponding to the IP version in the ethernet frame. These environment models were inferred through the context-aware environment refinement technique.
}
\label{listing:prvcheckoptions-harness}

\begin{minted}[
  fontsize=\footnotesize,
  linenos,
  breaklines,
  autogobble,
  escapeinside=||,
  frame=single,
  xleftmargin=1.5em,
  numbersep=5pt,
  style=colorful
]{c}

size_t uxIPHeaderSizePacket( const NetworkBufferDescriptor_t * nbuf )
{
    const EthernetHeader_t * pxEth = ( const EthernetHeader_t * ) nbuf->pucEthernetBuffer;
    if( pxEth->usFrameType == ( uint16_t ) ipIPv6_FRAME_TYPE )
    {
        return ipSIZE_OF_IPv6_HEADER;
    }
    return ipSIZE_OF_IPv4_HEADER;
}

void harness()
{
    size_t sock_len;
    |\textcolor{blue}{\_\_CPROVER\_assume}|(sock_len >= sizeof(FreeRTOS_Socket_t));
    FreeRTOS_Socket_t *sock = malloc(sock_len);
    |\textcolor{blue}{\_\_CPROVER\_assume}|(sock != NULL);

    size_t nbuf_len;
    |\textcolor{blue}{\_\_CPROVER\_assume}|(nbuf_len >= sizeof(NetworkBufferDescriptor_t));
    NetworkBufferDescriptor_t *nbuf = malloc(nbuf_len);
    |\textcolor{blue}{\_\_CPROVER\_assume}|(nbuf != NULL);

    size_t ethbuf_len;
    |\textcolor{blue}{\_\_CPROVER\_assume}|(ethbuf_len >= ipSIZE_OF_ETH_HEADER);
    nbuf->pucEthernetBuffer = malloc(ethbuf_len);
    |\textcolor{blue}{\_\_CPROVER\_assume}|(nbuf->pucEthernetBuffer != NULL);

    |\textcolor{blue}{\_\_CPROVER\_assume}|(nbuf->xDataLength <= ethbuf_len);

    const EthernetHeader_t *eth =
        (const EthernetHeader_t *) nbuf->pucEthernetBuffer;

    |\textcolor{blue}{\_\_CPROVER\_assume}|(
        ((eth->usFrameType == (uint16_t) ipIPv6_FRAME_TYPE) &&
         (nbuf->xDataLength >=
            (ipSIZE_OF_ETH_HEADER + ipSIZE_OF_IPv6_HEADER +
             ipSIZE_OF_TCP_HEADER))) ||
        ((eth->usFrameType != (uint16_t) ipIPv6_FRAME_TYPE) &&
         (nbuf->xDataLength >=
            (ipSIZE_OF_ETH_HEADER + ipSIZE_OF_IPv4_HEADER +
             ipSIZE_OF_TCP_HEADER))));

    |\textcolor{blue}{\_\_CPROVER\_assume}|(
        ipconfigIS_VALID_PROG_ADDRESS(sock->u.xTCP.pxHandleSent) == pdFALSE);

    prvCheckOptions(sock, nbuf);
}
\end{minted}
\end{listing}

\subsection{Case study 1: \texttt{prvCheckOptions}.}
\texttt{prvCheckOptions} processes TCP packet options using a socket pointer and a network-buffer pointer. 
\cref{listing:prvcheckoptions-harness} shows a unit proof, comprising its input and function model, created by \toolname.

\myparagraph{Scope.}
At scope level~1, \toolname verified the entry point and two reachable same-file callees, covering 107 lines of code.
The FreeRTOS proof verified only the 33-line entry point and excluded the callees, creating separate unit proofs to verify them.
Thus, \toolname maximized the number of properties verified with one unit proof, while FreeRTOS abstracted part of that behavior.

\myparagraph{Loop bounds.}
\texttt{prvCheckOptions} contains one loop that iterates over TCP packet options.
\toolname kept this loop at its initial bound of~2 because all memory-safety properties were covered and no in-scope memory access depended on further unwinding.
FreeRTOS used a bound of~41. 
It is not clear why they used such large loop bound even though the helper logic was already abstracted and no memory-safety access occur in or after that loop.
This shows how property-guided refinement can avoid unnecessary unwinding.

\myparagraph{Models.}
Both proofs used similar input models: valid struct pointers for input arguments (Lines 13--28) and preconditions that specify a lower bound on the size of received TCP packet (Lines 33--41).
\toolname also inferred the same packet-length precondition used by FreeRTOS, namely that the packet must contain the Ethernet, IP, and TCP headers.
For environment modeling, both proofs modeled \texttt{uxIPHeaderSizePacket} by returning the IP header size implied by the packet's IP version.
The main difference is that FreeRTOS also modeled the same-file helper \texttt{prvSingleStepTCPHeaderOptions}, while \toolname included that helper directly in scope.

Overall, this case shows that \toolname can replicate expert-like input and environment assumptions while checking more same-file behavior with a smaller loop bound.

\subsection{Case Study 2: \texttt{vDHCPProcess}}

\texttt{vDHCPProcess} function processes received DHCP packets and manages the DHCP state machine.

\myparagraph{Scope.}
\toolname includes \texttt{vDHCPProcess} and all reachable functions in its parent file, covering 17 functions and 507 lines of code.
The FreeRTOS proof includes only the 41-line entry function and excludes the reachable helper functions.
This difference reflects the goal of \toolname: rather than minimizing the proof to the entry point, it expands the scope to include code that may contain checkable memory-safety properties or may affect the entry point's memory-safety behavior.

\begin{listing}[t]
\centering
\caption{
    Loop-unwinding refinement proposed for a loop in \texttt{vProcessHandleOption}, reachable from vDHCPProcess, by the property-guided loop bound refinement technique. The technique identified a block of code which was not being verified, determined it resulted from insuffient loop bounds, and derived the minimal loop bound required for the loop exit condition to hold and the block of code to be verified.
}
\label{listing:vprocesshandleoption-unwind-refinement}

\begin{minted}[
  fontsize=\footnotesize,
  linenos,
  breaklines,
  autogobble,
  xleftmargin=1.5em,
  numbersep=5pt
]{json}
{
  "analysis": "The uncovered block starts at line 1095, immediately after the `while` loop at line 1088:
`while(uxTargetIndex < ipconfigENDPOINT_DNS_ADDRESS_COUNT)`.

  Root cause: `ipconfigENDPOINT_DNS_ADDRESS_COUNT` is 5 in this proof configuration (`test/unit-test/ConfigFiles/FreeRTOSIPConfig.h`), so the loop may require up to 5 iterations before exit...",
  
  "proposed_modifications": "Minimal change in Makefile only: add a loop-specific unwind bound for `vProcessHandleOption.1`.

Set:
`--unwindset vProcessHandleOption.1:6`
(5 needed iterations + 1 to allow falsifying the loop condition and exiting).",

  "updated_harness": ...
}
\end{minted}
\end{listing}

\myparagraph{Loop bounds.}
\toolname reaches eight loops but does not increase all bounds uniformly.
It keeps six loops at the default bound of 2, bounds \texttt{memcmp}'s internal loop at 7 because it compares six-byte MAC addresses, and increases the bound for \texttt{vProcessHandleOption.1} only after coverage feedback shows that the loop exit is needed to reach additional code.
\Cref{listing:vprocesshandleoption-unwind-refinement} shows this localized refinement.
In contrast, the FreeRTOS proof reaches only two loops and assigns both a uniform bound of 4.
This demonstrates the property-guided approach \toolname follows to derive appropriate loop bounds.

\myparagraph{Models.}
\toolname initializes the primitive argument as an unconstrained symbolic value and models the endpoint pointer as a valid allocated endpoint list.
It also infers, from verification feedback, that the global \texttt{pxNetworkEndPoints} must reference the input endpoint.
It also correctly infers, from a failed assertion in the program, that its \texttt{xSocketValid} model should return True after socket creation.

The FreeRTOS proof uses a similar endpoint model by passing the initialized global endpoint object directly to \texttt{vDHCPProcess}, but it additionally calls \texttt{prvCreateDHCPSocket} and \texttt{prvCloseDHCPSocket} around the target call, possibly to ensure a valid socket is available.
These differences identify the distinguishing feature of fidelity-oriented and safety-oriented unit proofs: fidelity-oriented proofs may use functions within the project to proactively set up valid calling context, while safety-oriented proofs recover these semantic knowledge as necessary and encode them as explicit assumptions in the unit proof.

\section{Instrumented Memory Safety Properties}
\label{app-sec:memory-safety-properties}

This appendix summarizes the memory-safety properties verified in this work as part of the experimental evaluation. These properties are instrumented by CBMC and genuine violations can represent high-impact security vulnerabilities.

\begin{enumerate}

  \item \textbf{Assertion}
  \begin{itemize}
    \item max allocation size exceeded: $\mathrm{malloc\_size} \leq \mathrm{MAX\_SIZE}$
    \item max allocation may fail: $\neg \mathrm{should\_malloc\_fail}$
  \end{itemize}

  \item \textbf{Pointer dereference}
  \begin{itemize}
    \item dereference failure: pointer NULL in $p$: $p \neq \mathrm{NULL}$
    \item dereference failure: pointer invalid in $p$: $\mathrm{valid}(p)$
    \item dereference failure: deallocated dynamic object in $p$: $\neg \mathrm{deallocated}(p)$
    \item dereference failure: dead object in $p$: $\neg \mathrm{dead}(p)$
    \item dereference failure: pointer outside object bounds in $p$: $\mathrm{offset}(p)+\mathrm{sizeof}(*p) \leq \mathrm{object\_size}(p)$
    \item dereference failure: invalid integer address in $p$: $\mathrm{object}(p)\neq\mathrm{object}(\mathrm{NULL}) \lor p=\mathrm{NULL}$
    \item no candidates for dereferenced function pointer: $\mathrm{false}$
    \item dereferenced function pointer must be $f$: $\mathrm{fp}=f$
    \item dereferenced function pointer must be one of $\{f_1,\ldots,f_n\}$: $\mathrm{fp}\in\{f_1,\ldots,f_n\}$
  \end{itemize}

  \item \textbf{Pointer arithmetic}
  \begin{itemize}
    \item pointer relation: pointer NULL in $p$: $p \neq \mathrm{NULL}$
    \item pointer relation: pointer invalid in $p$: $\mathrm{valid}(p)$
    \item pointer relation: deallocated dynamic object in $p$: $\neg \mathrm{deallocated}(p)$
    \item pointer relation: dead object in $p$: $\neg \mathrm{dead}(p)$
    \item pointer relation: pointer outside object bounds in $p+k$: $\mathrm{offset}(p)+k \leq \mathrm{object\_size}(p)$
    \item pointer relation: invalid integer address in $p+k$: $\mathrm{object}(p)\neq\mathrm{object}(\mathrm{NULL}) \lor p+k=\mathrm{NULL}$
  \end{itemize}

  \item \textbf{Pointer primitives}
  \begin{itemize}
    \item pointer invalid in $P(p,n)$: $p=\mathrm{NULL} \lor \mathrm{valid}(p)$
    \item deallocated dynamic object in $P(p,n)$: $p=\mathrm{NULL} \lor \neg \mathrm{deallocated}(p)$
    \item dead object in $P(p,n)$: $p=\mathrm{NULL} \lor \neg \mathrm{dead}(p)$
    \item pointer outside object bounds in $P(p,n)$: $p=\mathrm{NULL} \lor \mathrm{offset}(p)+n \leq \mathrm{object\_size}(p)$
  \end{itemize}

  \item \textbf{Array bounds}
  \begin{itemize}
    \item array lower bound in $a[i]$: $i \geq 0$
    \item array upper bound in $a[i]$: $i < |a|$
    \item array dynamic object upper bound in $p[i]$: $\mathrm{offset}(p)+i\cdot\mathrm{sizeof}(T) < \mathrm{object\_size}(p)$
    \item string constant lower bound in $s[i]$: $i \geq 0$
    \item string constant upper bound in $s[i]$: $i < |s|$
  \end{itemize}

  \item \textbf{Overflow}
  \begin{itemize}
    \item arithmetic overflow on signed $+$ in $x+y$: $\mathrm{MIN}\leq x+y\leq\mathrm{MAX}$
    \item arithmetic overflow on signed $-$ in $x-y$: $\mathrm{MIN}\leq x-y\leq\mathrm{MAX}$
    \item arithmetic overflow on signed $*$ in $x\cdot y$: $\mathrm{MIN}\leq x\cdot y\leq\mathrm{MAX}$
    \item arithmetic overflow on signed shl in $x \ll k$: $\mathrm{MIN}\leq x\cdot 2^k\leq\mathrm{MAX}$
    \item arithmetic overflow on signed unary minus in $-x$: $x \neq \mathrm{MIN}$
    \item arithmetic overflow on signed division in $x/y$: $\neg(x=\mathrm{MIN}\land y=-1)$
    \item result of signed mod is not representable in $x \bmod y$: $\neg(x=\mathrm{MIN}\land y=-1)$
  \end{itemize}

  \item \textbf{Undefined-shift}
  \begin{itemize}
    \item shift distance is negative in $x \ll k$: $k \geq 0$
    \item shift distance too large in $x \ll k$: $k < w$
    \item shift operand is negative in $x \ll k$: $x \geq 0$
  \end{itemize}

  \item \textbf{Division-by-zero}
  \begin{itemize}
    \item division by zero in $x/y$ or $x\bmod y$: $y \neq 0$
  \end{itemize}

  \item \textbf{Pointer}
  \begin{itemize}
    \item same object violation in pointer comparison: $\mathrm{object}(p)=\mathrm{object}(q)$
  \end{itemize}

  \item \textbf{Precondition instance}
  \begin{itemize}
    \item memcpy src/dst overlap: $\mathrm{object}(src)\neq\mathrm{object}(dst)\lor src+n\leq dst\lor dst+n\leq src$
    \item memcpy source region readable: $\mathrm{R\_OK}(src,n)$
    \item memcpy destination region writeable: $\mathrm{W\_OK}(dst,n)$
    \item memset destination region writeable: $\mathrm{W\_OK}(dst,n)$
    \item memmove source region readable: $\mathrm{R\_OK}(src,n)$
    \item memmove destination region writeable: $\mathrm{W\_OK}(dst,n)$
    \item strcpy/strncpy src/dst overlap: $\mathrm{object}(src)\neq\mathrm{object}(dst)\lor src+n\leq dst\lor dst+n\leq src$
    \item free argument must be NULL or valid pointer: $p=\mathrm{NULL}\lor \mathrm{R\_OK}(p,0)$
    \item free argument must be dynamic object: $p=\mathrm{NULL}\lor \mathrm{dynamic}(p)$
    \item free argument has offset zero: $p=\mathrm{NULL}\lor \mathrm{offset}(p)=0$
    \item double free: $p=\mathrm{NULL}\lor \neg\mathrm{deallocated}(p)$
    \item free called for new[] object: $p=\mathrm{NULL}\lor \neg\mathrm{new\_array}(p)$
    \item free called for stack-allocated object: $p=\mathrm{NULL}\lor \neg\mathrm{stack}(p)$
  \end{itemize}

  \item \textbf{Precondition}
  \begin{itemize}
    \item memcmp region readable: $\mathrm{R\_OK}(s,n)$
    \item memcpy region readable: $\mathrm{R\_OK}(s,n)$
    \item buffer nonnull: $buf\neq\mathrm{NULL}$
    \item size greater than zero: $size>0$
    \item buffer writable: $\mathrm{W\_OK}(buf,size)$
    \item object overlap: $\mathrm{object}(p)\neq\mathrm{object}(q)$
  \end{itemize}

\end{enumerate}

\section{Details of Reported Vulnerabilities}
\label{sec:new-vulns}

\begin{listing}[t]
    \small
  \centering
  \caption{Sample of out-of-bound write vulnerability on line 33 discovered by \toolname in RIOT-OS.}
  \label{listing:new-vuln}

    \begin{minted}[
    fontsize=\footnotesize,
    linenos,
    breaklines,
    xleftmargin=1.5em,
    numbersep=5pt,
    escapeinside=||,
    style=colorful
]{c}
// Public API. path is user-controlled.
ssize_t nanocoap_sock_post(nanocoap_sock_t *sock, const char       *path, ...)
{ return _sock_put_post(sock, path, ...);}

// nanocoap_sock_post -> _sock_put_post -> coap_opt_put_uri_pathquery -> coap_opt_put_string_with_len -> coap_put_option

size_t coap_opt_put_uri_pathquery(uint8_t *buf, uint16_t *lastonum, const char *uri) {
    coap_opt_put_string_with_len(buf, ..., string, ...);
}

// buf is a fixed size buffer from sock->hdr_buf
// const char *string contains user-provided string
size_t coap_opt_put_string_with_len(uint8_t *buf, ...,
                                    const char *string, ...)
{
    uint8_t *bufpos = buf;
    char *uripos = (char *)string;
    
    while (len) {
        uint8_t *part_start = (uint8_t *)uripos;
        bufpos += coap_put_option(bufpos, ..., part_start, ...);
    }
}

// odata is user-provided string. olen is its length.
size_t coap_put_option(uint8_t *buf, ..., const void *odata, size_t olen)
{
    assert(lastonum <= onum);

    n = _put_delta_optlen(buf, n, 0, olen);
    if (olen) {
        // Copy of olen bytes into fixed length buffer without validation.
        memcpy(buf + n, odata, olen);
        n += olen;
    }
    return (size_t)n;
}
\end{minted}
\end{listing}

\Cref{listing:new-vuln} illustrates a vulnerability in RIOT-OS (\texttt{nanocoap.c}) discovered by \toolname.
The bug is reached along the following execution flow.

\begin{itemize}
    \item \textbf{Initial harness.}
    \toolname first generates a harness for \texttt{coap\_opt\_put\_uri\_pathquery}, a root function in \texttt{nanocoap.c}.
    The harness initializes the \texttt{buf} and \texttt{string} arguments as nondeterministic buffers with unconstrained sizes.

    \item \textbf{Coverage refinement and bug exposure.}
    Stage~2 then resolves coverage gaps until all edges along the call path in line~5 of \cref{listing:new-vuln} are covered.
    At this point, verification exposes an out-of-bounds write at line~33.
    Although the code contains a loop at line~19, the bug manifests without increasing its unwinding limit.

    \item \textbf{Precondition generation.}
    In Stage~3, \toolname analyzes the counterexample trace and infers a precondition that constrains the input (buffer) length to eliminate the reported error.

    \item \textbf{Precondition validation.}
    The validator performs backward dataflow analysis to trace the constrained argument to its source.
    It determines that the relevant string originates from a parameter of the public API \texttt{nanocoap\_sock\_post}.
    Along this path, it finds no explicit constraint on the string length.
    It therefore flags the inferred precondition as violable in the real environment and reports the corresponding \texttt{memcpy} out-of-bounds write as a true vulnerability.
\end{itemize}

We reported this issue to the RIOT-OS maintainers, who acknowledged receipt and began an internal investigation.

\begin{table}
    \centering
    \caption{Comparison \toolname's performance when using open-source models}
    \begin{tabular}{l|cc}
    \toprule
    \textbf{Metric} & \textbf{GLM-5} & \textbf{MiniMax M2.5} \\
    \midrule
    Num Targets & 37.0 & 37.0 \\
    \midrule
    Compiles (\%) & 97.3\% & 75.7\% \\
    Structural validity (\%) & 81.1\% & 62.2\% \\
    Verification completes (\%) & 86.5\% & 73.0\% \\
    Generation succeeds (\%) & 78.4\% & 48.6\% \\
    \midrule
    Verification time (s) & 98.6 & 56.9 \\
    Avg component size (loc) & 99.9 & 54.6 \\
    Avg covered size (loc) & 90.9 & 42.7 \\
    Avg coverage (\%) & 96.1 & 81.5 \\
    \midrule
    Avg num. properties (\#) & 262.1 & 174.6 \\
    Avg num. verified properties (\#) & 253.0 & 170.3 \\
    Avg num. vulnerabilities (\#) & 2.1 & 1.2 \\
    \midrule
    Avg generation time (min) & 739.4 & 83.4 \\
    Avg API cost (\$) & 4.9 & 0.7 \\
    \midrule
    Avg proof size (loc) & 48.4 & 29.6 \\
    \bottomrule
    \end{tabular}
    \label{tab:rq3-model-comparison}
\end{table}

\section{Additional Evaluation Data and Results}
\label{sec:appendix:AdditionalData}

\cref{tab:evaluation-subjects} summarizes the evaluation subjects.
These are embedded operating systems developed and maintained by Google (Zephyr-RTOS), Amazon Web Services (FreeRTOS), and the international engineering community (RIOT-OS, Contiki-NG).

\begin{table}[t]
    \centering
    \caption{Summary statistics for the evaluation subjects.}
    \label{tab:evaluation-subjects}
    \small
    \begin{tabular}{llll}
    \toprule
        \textbf{Software} & \textbf{Size} & \textbf{Stars} & \textbf{Parent modules} \\
    \midrule
        Zephyr-RTOS & 2253971 & 15115 &
        BT, FS, USB, Shell, 6LoWPAN \\

        RIOT-OS & 417135 & 5718 &
        BT, FS, USB, Shell, 6LoWPAN \\

        Contiki-NG & 186095 & 1486 &
        BT, FS, USB, Shell, 6LoWPAN \\

        FreeRTOS & 836205 & 7280 &
        Shell, FS, OTA, MQTT, JSON \\
    \bottomrule
    \end{tabular}
\end{table}

Table~\ref{tab:rq3-model-comparison} compares \toolname's performance when using two open-source models, \textit{GLM-5} and \textit{MiniMax M2.5}, across 37 targets.

\end{document}